\documentclass{aa}

\usepackage{txfonts}

\usepackage{natbib}
\usepackage{float}
\usepackage{color}

\usepackage{graphicx}   % Including figure files
\usepackage{amssymb}    % Extra maths symbols

\usepackage{caption,subcaption}

\usepackage{booktabs, multirow}
\usepackage{comment}
\usepackage{amsmath,siunitx}
\usepackage{soul}% for underlines
\usepackage[table]{xcolor} % for cell colors
\usepackage{changepage,threeparttable} % for wide tables
\usepackage[utf8]{inputenc}
\usepackage[T1]{fontenc}
\usepackage{tablefootnote}
\usepackage{longtable}
\usepackage{footnote}
\usepackage{placeins}
\usepackage{txfonts}
\usepackage[]{hyperref}
\hypersetup{
  colorlinks,
  citecolor=blue,
  linkcolor=red,
  urlcolor=blue}
\usepackage{afterpage}
\usepackage{newtxtext,newtxmath}
\bibpunct{(}{)}{;}{a}{}{,} % to follow the A&A style

\DeclareRobustCommand{\VAN}[3]{#2}
\let\VANthebibliography\thebibliography
\def\thebibliography{\DeclareRobustCommand{\VAN}[3]{##3}\VANthebibliography}

\renewcommand\thesubfigure{(\alph{subfigure})}
\newcommand{\TEMPO}{\texttt{TEMPO}}
\newcommand{\TEMPOTWO}{\texttt{TEMPO2}}

\newcommand{\msun}{$\rm M_{\sun}$}

\newcommand{\chisquare}{\chi^2}

\newcommand{\Pbdot}{\dot{P}_{\rm b}}
\newcommand{\alphap}{\alpha_{\rm p}}
\newcommand{\alphac}{\alpha_{\rm c}}
\newcommand{\Pbdotobs}{\dot{P}_{\rm b \rm ,obs}}
\newcommand{\Pbdotint}{\dot{P}_{\rm b \rm ,int}}
\newcommand{\Pbdotshk}{\dot{P}_{\rm b \rm ,shk}}
\newcommand{\Pbdotgal}{\dot{P}_{\rm b \rm ,gal}}

\newcommand{\Pbdotgr}{\dot{P}_{\rm b \rm ,GR}}
\newcommand{\Pbdotx}{\dot{P}_{\rm b \rm ,x}}
\newcommand{\Pbdotxs}{\dot{P}_{\rm b \rm ,xs}}

\newcommand{\Pdotobs}{\dot{P}_{\rm obs}}
\newcommand{\Pdotint}{\dot{P}_{\rm int}}

\newcommand{\Pb}{P_{\rm b}}

\newcommand{\Mp}{M_{\rm p}}
\newcommand{\Mc}{M_{\rm c}}

\newcommand{\cosi}{\rm cos \, \rm {i}}

\begin{document}

\title{Relativistic effects in a mildly recycled pulsar binary: PSR~J1952+2630}
\titlerunning{Relativistic effects in PSR~J1952+2630}

\author{T.~Gautam \inst{1} \thanks{Member of the International Max Planck Research School (IMPRS) for Astronomy and Astrophysics at the University of Bonn} \and  P.~C.~C.~Freire\inst{1} \and A. Batrakov\inst{1} \and M. Kramer\inst{1} \and C. C. Miao \inst{2} \and E. Parent\inst{3}\inst{4} \and  W. W. Zhu\inst{2}}
\institute{Max-Planck-Institut f\"{u}r Radioastronomie, Auf dem H\"{u}gel 69, D-53121 Bonn, Germany\label{1} \and 
National Astronomical Observatories, Chinese Academy of Sciences, Beijing 100101, China\label{2} \and
Institute of Space Sciences (ICE, CSIC), Campus UAB, Carrer de Can Magrans s/n, E-08193, Barcelona, Spain\label{3} \and
Institut d'Estudis Espacials de Catalunya (IEEC), Carrer Gran Capità 2-4, E-08034 Barcelona, Spain\label{4} 
\\
\email{tgautam@mpifr-bonn.mpg.de}}

\date{}

\abstract
  {We report the results of timing observations of PSR J1952+2630, a 20.7 ms pulsar in orbit with a massive white dwarf companion. We performed six months of timing observations with the Arecibo radio telescope in 2020 and used data from FAST from 2021. Together with previously published data, this represents a total timing baseline of 11 years since the discovery of the pulsar in 2010.
  For the first time, we present a polarimetric profile of the pulsar and determine its rotation measure (RM), $-145.79 \pm 0.15 \, \rm rad \, \rm m^{-2}$. With the increased timing baseline, we obtain improved estimates for astrometric, spin, and binary parameters for this system. In particular, we obtain an improvement of an order of magnitude on the proper motion, and, for the first time, we detect three post-Keplerian parameters in this system: the advance of periastron $\dot{\omega}$, the orbital decay $\dot{P}_{\rm b}$, and the Shapiro delay (measured in the form of the $h_3$ parameter). With the detection of these relativistic effects, we constrain the pulsar mass to 1.20$^{+0.28}_{-0.29}$ \msun\, and the mass of its companion to 0.97$^{+0.16}_{-0.13}$ \msun. The current value of $\Pbdot$ is consistent with the General Relativity expectation for the masses obtained using $\dot{\omega}$ and $h_3$. The excess ($4.2 ^{+70.2}_{-73.1} \rm fs \, s^{-1}$) represents a limit on the emission of dipolar gravitational waves (GWs) from this system. This results in a limit on the difference in effective scalar couplings for the pulsar and companion (predicted by scalar-tensor theories of gravity; STTs) of $|\alphap-\alphac| < 4.8 \times 10 ^{-3}$ (68\% C.L.), which does not yield a competitive test for STTs. However, our simulations of future timing campaigns of this system, based on the timing precision we have achieved with FAST, show that by 2032, the precision of $\Pbdot$ and $\dot{\omega}$ will allow for much more precise masses and much tighter constraints on the orbital decay contribution from dipolar GWs, resulting in $|\alphap-\alphac| < 1.3 \times 10 ^{-3}$ (68\% C.L.). For comparison, we obtain $|\alphap-\alphac| < 1.9 \times 10 ^{-3}$ and $ < 3.3 \times 10 ^{-3}$ from PSR~J1738+0333 and PSR~J2222$-$0137, respectively. We also present the constraints this system will place  on the $\{\alpha_0, \beta_0\}$ parameters of Damour-Esposito-Far{\`e}se (DEF) gravity by 2032. They are comparable to those of PSR~J1738+0333. Unlike PSR~J1738+0333, PSR~J1952+2630 will not be limited in its mass measurement and has the potential to place even more restrictive limits on DEF gravity in the future. Further improvements to this test will likely be limited by uncertainties in the kinematic contributions to $\dot{P}_{\rm b}$ because of the lack of precise distance measurements.}


\maketitle 
%%%%%%%%%%%%%%%%%%%%%%%%%%%%%%%%%%%%%%%%%%%%%%%%%%

%%%%%%%%%%%%%%%%% BODY OF PAPER %%%%%%%%%%%%%%%%%%

\section{Introduction}

Millisecond pulsars (MSPs), especially those located in binary systems, provide us with a unique opportunity to gain insights in fundamental physics, not only for tests of gravity theories \citep{2015CQGra..32x3001B}, but also for the study of the unknown state of matter (and its equation of state, or EOS) in the centres of neutron stars (NSs), which is a fundamental problem in the study of the strong nuclear force \citep{2016ARA&A..54..401O}. These studies are performed using a technique known as pulsar timing, which is a method of modelling the physical effects affecting the times of arrival (ToAs) of their radio pulsations ( \citealt{1992RSPTA.341..117T}). These are the spin and its variations, astrometry, and interstellar effects. This technique is especially powerful for MSPs because of their stable rotation and the high ToA precision. Furthermore, because of their evolution, most MSPs are members of binary systems. In these cases, their timing in addition provides exquisite measurements of their orbital motion. The detection of relativistic effects in the orbital motion provided the first confirmation of the existence of gravitational waves \citep{1989ApJ...345..434T} and today allows a wide range of studies of gravity theories (e.g. \citealt{2012MNRAS.423.3328F,2013Sci...340..448A,2019CQGra..36v5009A,2020A&A...638A..24V,2021PhRvX..11d1050K,2021A&A...654A..16G,2022CQGra..39kLT01Z}) and constraints on the EOS (e.g. \citealt{2016ARA&A..54..401O,2021ApJ...915L..12F}).

Thus, it is important to discover and time additional pulsar systems where orbital relativistic effects can be precisely measured. 
This was one of the main goals of the Pulsar Arecibo L-band Feed Array (PALFA) survey \citep{2006ApJ...637..446C,2015ApJ...812...81L}, which  was carried out from 2004 to 2020 using the seven-beam receiver of the 305 m Arecibo radio telescope located near Arecibo, Puerto Rico, USA. This scanned the Galactic longitudes accessible to that telescope (especially $25^\circ < \ell < 77^\circ$) for low Galactic latitudes ($| b | < 5^\circ $).
The high sensitivity of the telescope enabled the survey to achieve high sensitivity with relatively short pointings. This made it sensitive to compact systems with very high accelerations, as shown by the discovery of three compact double neutron star systems (DNSs): PSR~J1906+0746, the youngest DNS \citep{2006ApJ...640..428L,2015ApJ...798..118V}, PSR~J1913+1102, the first merging, strongly asymmetric DNS \citep{2016ApJ...831..150L,2020Natur.583..211F}, and PSR~J1946+2052, the most compact DNS known in our Galaxy, with an orbital period of only 1h 53m \citep{2018ApJ...854L..22S}. 
Overall, the survey discovered a total of 207 pulsars\footnote{\url{http://www.naic.edu/~palfa/newpulsars/}}, 38 of which are MSPs, defined here somewhat arbitrarily as having $P < 25$ ms.

In this survey, a computationally intensive analysis was carried out with the help of distributed computing project, Einstein@Home \citep{2010Sci...329.1305K,2013ApJ...773...91A} to search for isolated or binary pulsars with orbits longer than 11 minutes. PSR~J1952+2630 was the first binary pulsar system discovered with this pipeline \citep{2011ApJ...732L...1K}.
With a spin period of 20.7 ms, it is one of the aforementioned MSPs. It has a relatively short 9.4 hr orbit with a massive white dwarf (WD) companion. The orbital eccentricity was already known to be low ($e \leq$ $1.7\times 10^{-3}$).
Assuming a pulsar mass ($M_{\rm p}$) of $1.4 \,$\msun, they obtained a minimum $\Mc$ of 0.95 \msun.
The system's low eccentricity excludes the possibility that the companion might be another NS; it is very likely a massive WD star.
The relatively long spin period (compared to most MSPs) is commonly found among other similar intermediate-mass binary pulsar systems (IMBPs; \citealt{2001ApJ...548L.187C}); the difference is that the orbit is unusually short for such systems.

Unlike the commonly accepted formation scenario of low-mass binary pulsars (LMBPs) from low-mass X-ray binaries (LMXB), where an NS spins up or is recycled during the accretion of matter from its evolved low-mass companion star \citep{1982Natur.300..728A,1982CSci...51.1096R,1991PhR...203....1B,1993ARA&A..31...93V}, intermediate-mass X-ray binaries (IMXBs) are thought to undergo a completely different recycling mechanism. Their hydrogen-rich massive donor stars (2-10 \msun) do not provide sufficient time to recycle the NS through accretion. A possible formation mechanism of IMBPs in short orbits with a carbon-oxygen (CO) WD companion was discussed by \cite{2011MNRAS.416.2130T,2012MNRAS.425.1601T}. It involves a common-envelope (CE) phase that circularizes the binary and reduces its orbital period, followed by a case BB Roche-lobe overflow (RLO) phase, which continues long enough to recycle the NS to its short rotational periods. 

To develop a phase-connected timing solution and thus derive precise spin, astrometric, and orbital parameters, \cite{2014MNRAS.437.1485L} carried out further timing observations of PSR~J1952+2630 with the Arecibo telescope. With their added baseline, they confirmed the high mass of the companion $>$ 0.93\msun \ and measured a low but significant orbital eccentricity, $e \sim 4.1 \times 10^{-5}$.
The measured spin-down implies that the characteristic age of the system is 77 Myr, younger than most recycled pulsars, especially when compared with other IMBPs.
 %Not necessary: Thus, assuming the current kinematic properties of this system, \citealt{2014MNRAS.437.1485L} modelled the mass-transfer process from a He-star, predicting that it first underwent a common envelope (CE) phase which did not continue long enough, and then a case BB Roche Lobe Overflow (RLO) episode that provided the pulsar with its current small periodicity. 
To investigate the nature of the WD progenitor, they simulated evolutionary models using the Langer stellar evolution code \citep{2011MNRAS.416.2130T,2012MNRAS.425.1601T} and predicted two solutions for the companion progenitor's nature: a) a 1.17 \msun ONeMg WD created from a 2.2 \msun He star, and b) a 1.02 \msun CO WD formed from a 1.9 \msun He star. Because the orbital solution at that time did not allow a measurement of the PK parameters, there was no measurement of the component masses. Hence, neither of these scenarios could be confirmed. However, they suggested that continued timing would eventually lead to a measurement of three PK parameters, $\dot{\omega}$, $\Pbdot,$ and possibly the orthometric amplitude of the Shapiro delay $h_3$ \citep{2010MNRAS.409..199F}, indicating that additional timing could result in precise mass measurements and in a clarification of the real nature of this system.

This work presents the results from the radio timing of this pulsar, which make use of an additional six months of timing observations performed with the Arecibo radio telescope in 2020 and two additional observations carried out with the Five hundred meter aperture spherical
radio telescope (FAST) \citep{2011IJMPD..20..989N,2020Innov...100053Q} in 2021. With an 11-year-long timing baseline, we can now measure the proper motion of the system, as predicted, and $\dot{\omega}$, $\Pbdot$, and $h_3$.

The structure of the paper is as follows.  Section~\ref{sec:observations} gives details on all observations and how the resulting data were analysed. In Section~\ref{sec:results}, we present the main results of our polarimetry and timing analysis, with a discussion of the main new parameters. In Section~\ref{sec:measurements}, we discuss the masses for the components, and the constraints from the orbital decay measurement of this system. In Section~\ref{sec:lab}, we discuss the potential of PSR~J1952+2630 as a gravitational laboratory. Here we simulate future timing in order to determine the level of precision we might achieve in our mass measurements and tests of gravity theories with this system, in particular, limits on dipolar gravitational-wave (GW) emission, a prediction of alternative theories of gravity. Using our improved proper motion measurement, we can now calculate the uncertainty on the kinematic contributions to $\Pbdot$, which provides the ultimate limit on the precision of intrinsic $\Pbdot$. In this section, we also discuss the constraints this system will place on Damour-Esposito-Far{\`e}se (DEF) gravity. We list our conclusions in Section~\ref{sec:discussion}.

\section{Observations and data reduction} 
\label{sec:observations}

\subsection{Observations}

After the discovery of PSR~J1952+2630 in 2010, the binary was regularly observed from July 2010 to September 2011 as part of the PALFA survey with the seven-beam ALFA receiver of the Arecibo telescope. These observations were taken with the Mock Spectrometers\footnote{\url{https://www.naic.edu/~astro/mock.shtml}} (see details in Table~\ref{table:observations}).
Afterwards, \cite{2014MNRAS.437.1485L} carried out dedicated timing observations of this system using the L-wide receiver of the Arecibo telescope, which has a wider bandwidth; see details in Table~\ref{table:data_parameters}. In the beginning, three 3hr observations were performed, and subsequently, 1hr observations were carried out every month for the next 11 months. These observations were taken in Stokes I, that is, only total intensity was recorded with integrated polarisation information. These data were later dedispersed and folded at the spin period of the pulsar (according to the best available ephemeris), producing an archive of pulse profiles (henceforth "archive") with 512 profile bins, as described by \cite{2014MNRAS.437.1485L}.

From February 2020 to July 2020, we performed another timing campaign on this system with Arecibo, again using the L-wide receiver. This campaign made use of the new more sensitive wide-bandwidth backend system, the Puerto Rico Ultimate Pulsar Processing Instrument (\texttt{PUPPI}\footnote{\url{http://outreach.naic.edu/ao/scientist-user-portal/astronomy/puppi-observing-and-support-guide}}). Importantly, and unlike the Mock spectrometer, this backend can use coherent dedispersion to completely remove any signal smearing caused by the dispersion; this is especially important for PSR~J1952+2630 given its high DM of $\sim \, 315.3\, \rm cm^{-3} \, pc$; furthermore, the backend allows the recording of full-Stokes data. We observed this system for a total of 24 hr: 12 observations of 2 hr each. %, with each sample recorded every 0.64 $\mu$s. 
     These data were folded by PUPPI using the best available ephemeris for the pulsar in 2048-bin profiles, directly producing pulse profile archives. This greatly reduces the amount of data to be recorded, thus allowing the full-Stokes data to be recorded as well.

Additionally, to follow up on this system with better sensitivity, regular timing observations are being performed with the FAST radio telescope. In this work, we include the data from the latest two FAST observations in 2021. These data were taken with the ROACH backend in full Stokes. High-resolution search-mode data with 4096 channels were recorded originally, but for our purpose of determining achievable timing precision with FAST for simulations, we only used its downsampled version (see details in Table~\ref{table:data_parameters}). 
% Table \ref{table:observations} summarizes all the observations used in this work.

\begin{table*}
\caption[]{Observation details}
\label{table:observations}
\footnotesize
\centering
\renewcommand{\arraystretch}{0.5}
\setlength{\tabcolsep}{5.0pt}
\vskip 0.1cm
\begin{tabular}{lcccccccccc}
\hline
\hline
 {}&  &  & &  &  & & \\
{Observatory} & {Receiver} & {Backend} & {Epoch} & Date & {Coherently} \\ {}& & & {(MJD)} & &  {de-dispersed} \\ \midrule
{Arecibo}& {L-wide} & Mock  & {55407-55816} & July 2010 - Sept 2011 &  no  \\
 {}&  &  spectrometers & &  \\
 {}&  &  & & &  \\
{Arecibo}&{L-wide} & Mock &{55889-56218} & Nov 2011 - Oct 2012 &  no \\
 {}&  &  spectrometers & & \\
  {}&  &  & & &  \\
{Arecibo}&{L-wide}&PUPPI &{58881-59053} & Feb - July 2020 &  yes \\ %numchan=512
 {}&  &  & &  &  & & \\
{FAST}&{19-beam} & ROACH & {59507, 59554} & Oct - Dec 2021 &  no \\%numchan=256
\hline
\hline
\end{tabular}
\end{table*}

\begin{table*}
\caption[]{Data parameters.}
\label{table:data_parameters}
\footnotesize
\centering
\renewcommand{\arraystretch}{0.5}
\setlength{\tabcolsep}{5.0pt}
\vskip 0.1cm
\begin{tabular}{lcccccccccc}
\hline
\hline
 {}&  &  & &  &  & & \\
{Epoch}  & {Central} & {Bandwidth} & {Number of}& {Sampling} & {Number of} & {Typical} \\ {(MJD)} & {frequency (MHz)} &  {(MHz)}& channels &  time ($\mu \rm s$) & profile bins & length (min) \\ \midrule
 {55407-55816}& 1300.168 & 172.032 & 512 & 65.47 & 512 & 5-10 \\
  &   &    &  & &  \\
 &1450.168 & 172.032 & 512 & 65.47 & 512 & 5-10  \\
 %numchan=512
  &    & & & & \\
 {55889-56218} &  1185.958 & 172.032 & 2048 & 83.33 & 512 & 60  \\
  &   &    &  &  & \\
  &1358.042 & 172.032 & 2048 & 83.33 & 512 & 60 \\
  &   & & & \\
  &   1530.042 & 172.032 & 2048 & 83.33 & 512 & 60 \\
  &    &  & & \\
  &   1702.042 & 172.032 & 2048 & 83.33 & 512 & 60 \\ %numchan=2048
   &    & & & \\
{58881-59053} & 1380.781 & 800 & 512 & 0.64 & 2048 & 120 \\ %numchan=512
 &    & & & & \\
 {59507, 59554}  & 1250.0 & 500 & 256 & 49.152 & 256 & 15 \\%numchan=256
\hline
\hline
\end{tabular}
\end{table*}

\subsection{Data reduction}
\label{sec:datared}

To remove the temporal and spectral radio interference (RFI) created by terrestrial signals in the PUPPI archives, we used the \texttt{CoastGuard} \citep{2016MNRAS.458..868L} RFI-removal script, \texttt{clean.py}. We made use of two of its algorithms: \texttt{rcvrstd} and \texttt{surgical}, the former zero-weights the bad frequency channels that have no receiver response, and the latter removes the folded profiles affected by RFI. This is done by first fitting individual sub-integration profiles with the integrated pulse profile, and then removing the sub-integrations that were outliers in the resulting residual profiles (see \citealt{2016MNRAS.458..868L} for more details of the algorithms).

To achieve consistent results and improve post-fit residuals, we re-analysed the
archives obtained from the Mock spectrometers by \cite{2014MNRAS.437.1485L} in a way consistent
with our analysis of the PUPPI archives. In this analysis we used the standard routines in the \texttt{PSRCHIVE} pulsar package \citep{2004PASA...21..302H,2012AR&T....9..237V}. Each of the initial PSRFITS format files were first cleaned of RFI using the \texttt{PSRCHIVE} interactive cleaning library \texttt{pazi}.

For each Mock spectrometer archive, the pulse profiles were added in time and frequency (i.e. they were ``scrunched'') using the \texttt{PSRCHIVE} routine \texttt{pam}. This resulted in much smaller archives with two sub-banded profiles per epoch (of typically 5-10 min) at frequencies of 1300 MHz and 1400 MHz for earlier data (2010-2011), and four sub-banded profiles (of typically 15 minutes) at frequencies mentioned in Table~\ref{table:observations} for the data taken in 2011-2012. For the latest PUPPI data, we scrunched the 11 min archive files into eight sub-bands and a single 11-min sub-integration using \texttt{pam}; and for the FAST data, we created 30 sub-integrations for every 2-minute file.

For the Mock archives, we then phase-aligned the profiles at their corresponding central frequencies using an improved ephemeris and added them together using \texttt{psradd}. We created two standard profiles at frequencies 1300 MHz and 1530 MHz. Both the standard profiles were then fully scrunched in time and frequency and were fitted with several von Mises functions using \texttt{paas} to create two analytic templates. Of these, the 1300 MHz template was cross-correlated to create ToAs using \texttt{pat} for 1300 MHz, 1185 MHz, and 1358 MHz, while the 1530 MHz template was used for 1450 MHz, 1530 MHz, and 1702 MHz archive files.
Because all these data are broadband, pulse profile variations across the observed frequencies can increase the post-fit uncertainties. We therefore used standard templates with multiple frequencies.

To determine the best scrunching scheme for PUPPI data, we created ToAs using both the frequency-integrated and multi-frequency (sub-banded) templates (created using the FDM algorithm of \texttt{pat}). Additionally, sharp features in the polarisation profile can improve the timing precision significantly. The two templates were therefore created with and without integrated polarisation (using matrix template matching, the \texttt{pat} MTM algorithm). We then compared the resulting fit of the timing model using the four templates mentioned above. The sub-banded template with integrated polarisation had the least post-fit root-mean-square (RMS) and parameter uncertainties, therefore we used it to create the final ToAs. All the PUPPI observations were added together to create a single multi-frequency template with eight sub-bands, which was then cross-correlated with the similarly sub-banded data using \texttt{pat}. We obtain a typical ToA residual uncertainty of 20 $\mu$s.

To create ToAs from two FAST observations, we integrated the observations to create a single archive with integrated frequency channels and then used \texttt{paas} to create a noiseless standard template.

\subsection{Timing analysis and binary models}

The timing analysis was performed using the \TEMPO\footnote{\url{http://tempo.sourceforge.net}} pulsar timing software. In this analysis, the ToAs are first converted into a terrestrial time standard, UTC(NIST), which is a version of Coordinated Universal Time maintained by the US National Institute of Standards and Technology. The converted time was then translated into the Solar System barycentric (SSB) time using the Solar System ephemeris, DE436 (maintained by NASA's Jet Propulsion Laboratory, JPL; \citealt{Folkner_Park+2016}).
The program then minimizes the difference between the ToAs and the arrival times predicted by the model ephemeris, known as the timing residuals, by varying the timing parameters. These timing parameters are presented in barycentric dynamical time (TDB) units. As a starting point, we used the ephemeris from \cite{2014MNRAS.437.1485L}.

The descriptions of the orbit we used are based on the DD model of \cite{1986AIHS...44..263D}. This provides a quasi-Keplerian description of the orbit, with small perturbations, which in this case are assumed to be relativistic, that can be detected in the timing. These are quantified in a theory-independent way by a set of post-Keplerian (PK) parameters. This description was found to work for many alternative theories of gravity \citep{PhysRevD.45.1840}. One of these PK parameters, measurable in binary systems with eccentric orbits (where we can measure the longitude of periastron, $\omega$), is the rate of advance of periastron, $\dot{\omega}$.
% A second PK parameter, the Einstein delay ($\gamma$), is a combination of the varying special relativistic time dilation and the gravitational redshift a pulsar will experience at the different phases of an eccentric orbit; it is generally only measurable when the timing baseline covers significant changes of $\omega$, otherwise it is impossible to separate it from the geometric delay caused by the orbital motion (but there are exceptions, see detailed discussion by \citealt{2019MNRAS.490.3860R}).
A second PK parameter, the orbital decay caused by the emission of GWs, $\Pbdot$, can be measured in binaries with short orbits due to their higher rates of GW damping. Additionally, in binaries with a favourable combination of high orbital inclination ($i$), high companion mass ($\Mc$), and good timing precision, we can measure an additional relativistic effect known as Shapiro delay \citep{1964PhRvL..13..789S}, which is a retardation of the arrival times of the radio pulses caused by the spacetime curvature near its companion; this is quantified by two PK parameters that yield direct estimates of $\Mc$ and $\sin i$ in a wide range of gravity theories. This effect is especially pronounced near superior conjunction, when the companion is closest to our line of sight to the pulsar. A final PK parameter, the Einstein delay, is not measurable for systems with low orbital eccentricities such as PSR~J1952+2630. We therefore do not discuss it here.

The measurement of PK parameters is important: two such parameters in the same system can generally constrain the pulsar mass ($\Mp$) and $\Mc$. To do this, we must be confident that the observed orbital effects are purely relativistic, and then assume a particular gravity theory, such as general relativity (GR), to calculate the masses from the PK parameters. Because each EOS for dense matter predicts a maximum NS mass, the measurement of a high NS mass can rule out many acceptable EOSs \citep{2013Sci...340..448A,2021ApJ...915L..12F}. Measurement of three or more PK parameters in the same system provides a test of the self-consistency for any gravity theories that can link the two masses and Keplerian orbital parameters to the PK parameters (see \citealt{2021PhRvX..11d1050K} for the ultimate example of this). 

In this work, the orbital parameters of the system are measured using two different binary models based on the DD description: the DDGR and the ELL1H+ models.
In addition to the five Keplerian parameters (projection of the semi-major axis \textit{x}, eccentricity \textit{e}, orbital period $\Pb$, and both the epoch ($T_0$) and longitude of periastron ($\omega$) at $T_0$), the DDGR model fits directly for $\Mc$ and total mass ($M_{\rm T}$) by assuming that all relativistic effects that can be detected in the TOAs are as predicted by GR for that pair of masses.

The ELL1H+ model is, like the DD model, theory-independent. We used it to investigate which relativistic parameters can be detected in the timing, and also to provide a more accurate description of some of the orbital parameters.
For binaries with low eccentricity, the Keplerian parameters $T_0$ and $\omega$ that are used in the DD model are strongly correlated. This leads to high uncertainties in their measurements. To avoid this strong correlation, the ELL1 model \citep{2001MNRAS.326..274L} takes the epoch of ascending node, $T_{\rm asc}$ , as reference rather than $T_0$; this quantity is well defined even for circular orbits. Additionally, it fits for the Laplace-Lagrange parameters $\epsilon_1 = e\, \rm sin \omega$ and $\epsilon_2 = e\, \rm cos \omega$ instead of \textit{e} and $\omega$. Furthermore, by using $T_{\rm asc}$ as the reference for measuring ${P}_{\rm B}$, we also avoid its strong correlation with the measurement of $\dot{\omega}$. Because of the this lack of correlation, we were able to measure $T_{\rm asc}$ with a precision of $\sim$ 20100 times better than $T_0$, and obtained an improvement on $P_{\rm b}$ measurement by a factor of 3200.

Another strong correlation especially for low orbital inclinations is seen in the Shapiro delay implementation in the DD and ELL1 models, where it is quantified by two PK parameters, range ($r \equiv T_\odot \Mc$, where $T_\odot \equiv {\cal G} {\cal M}_{\odot}^{\rm N} / c^3 = 4.9254909476412669... \mu$s is an exact quantity; see \citealt{2016AJ....152...41P}) and shape ($s \equiv \, \sin i$). 
The ELL1H model, an extension of the ELL1 model, prevents this correlation by re-parametrising the Shapiro delay with two different PK parameters, the orthometric amplitude ($h_3$) and orthometric ratio ($\varsigma$) \citep{2010MNRAS.409..199F}. In our analysis, we used the ELL1H+ model (implemented in \TEMPO by N. Wex for the analysis of the PSR~J2222$-$0137 data; see \citealt{2021A&A...654A..16G}), which makes use of the exact expression of Eq. (31) of \cite{2010MNRAS.409..199F}, and includes an additional term of the order of $xe^2$ in the R\"omer delay expansion (as shown in Eq. 1 of \citealt{2019MNRAS.482.3249Z}).

\section{Results}
\label{sec:results}
\subsection{Polarisation calibration}

\begin{figure}
\centering
        \includegraphics[width=1.0\linewidth,trim={1.5cm 1.5cm 1.5cm 1.5cm},clip]{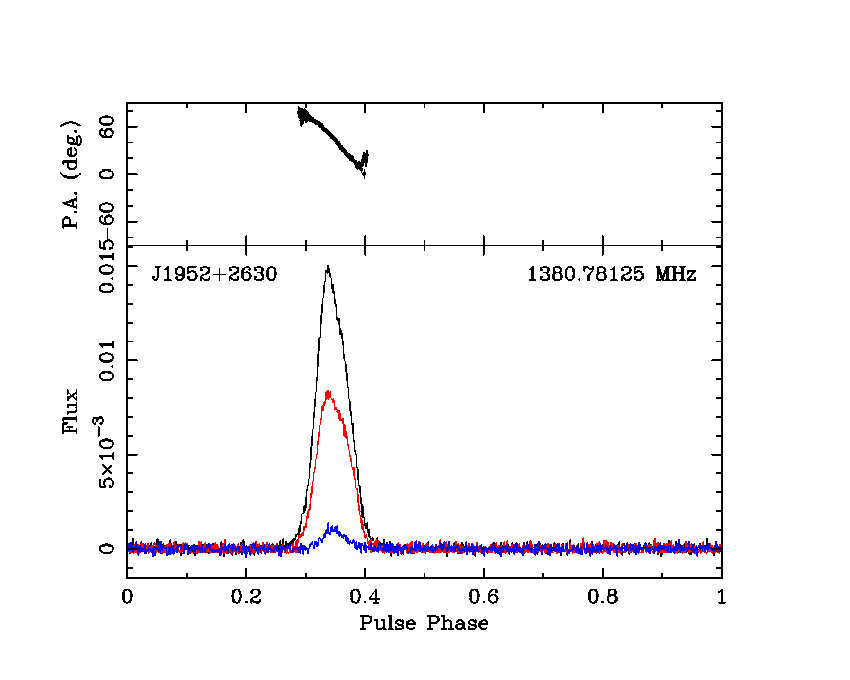}
        \caption{Total intensity (black), and linear and circular polarisation (red and blue, respectively) as a function of longitude. The top panel shows the polarisation angle (PA) swing across the rotational phase.}
        \label{fig:pol-profile}
\end{figure}

Because pulsars are one of the most strongly polarised radio sources, an analysis of polarisation in pulsars helps understand their radio beam geometry and its emission mechanism \citep{1969ApL.....3..225R}. We performed the polarisation calibration on the PUPPI data set of this pulsar using noise-diode scans that were carried out for $\sim$ 92 seconds at the beginning of each observation. Each calibration scan and its corresponding target observation was manually cleaned with \texttt{pazi}, and then calibration was applied on each epoch using \texttt{pac} routine of \texttt{PSRCHIVE}. To determine the best-fit rotation measure (RM), we first created an integrated profile by adding all the observations in time using \texttt{psradd} and \texttt{pam}, and then fit for RM in this high S/N profile (which had 512 channels and 2048 sample bins) using \texttt{rmfit} routine. This routine runs trials with various RM values to find the best fit by optimising the fraction of linear polarisation in the profile. We deduced an RM of $-$145.79$\pm$0.15 rad m$^{-2}$ using this method. Figure~\ref{fig:pol-profile} shows the calibrated polarisation profile of the pulsar with both linear and circularly polarised components. To correct for the Faraday rotation, all the observations were then de-rotated using the above-mentioned RM value to create archive files with better profile resolution. These archives were then used to create a polarised template that was used to decide the best ToA scheme. 

The polarisation position angle (PA) sweep across the rotational phase of the pulse can allow us to predict the geometry of the pulsar beam. According to \cite{1969ApL.....3..225R}, the PA sweep creates an S-shaped curve along with the pulse phase that can be modelled using the rotating vector model (RVM), given by
\begin{equation}
    \rm tan(\psi-\psi_0)= \frac{\rm sin (\alpha) \, \rm sin(\phi-\phi_0)}{\rm sin(\alpha + \beta) \rm cos(\alpha) - \rm cos(\alpha + \beta) \rm sin(\alpha) \rm cos(\phi - \phi_0)} ,
\end{equation}
where $\psi$ is the observed PA of the pulsar beam as it sweeps across our line of sight, $\alpha$ is the projection of the magnetic inclination angle, and $\beta$ is the impact parameter or closest approach of the magnetic axis to the line of sight. $\, \psi_0$ and $\alpha_0$ represent their respective values at a fiducial reference plane. 
We performed an MCMC fit of this model on the PA swing of its profile, which provided a constraint on $\alpha$ of $19.5^{+8.4}_{-8.5}$ deg and viewing angle, $\zeta$ (defined as $\alpha + \beta$) of $27.5^{+11.6}_{-11.8}$ deg. $\zeta$ represents the value for inclination angle ($i$), if the spin vector of the pulsar is aligned with the orbital angular momentum vector. However, because the duty cycle of the pulsar is small ($\sim$ 7$\%$) in this case, a large covariance between $\alpha$ and $\zeta$ exists because of which we cannot constrain \textit{i} and $\beta$ well. The minimum inclination of the binary calculated from its mass function and the allowed maximum $\Mc$ of 1.48 \msun \, is 28 deg, while the inclination derived from timing analysis is 71.72 deg. Thus, the system inclination was only poorly constrained from the above-mentioned RVM fit. 

%$\psi_0$ = 36.38  ( +0.66, -0.61 )
%$\Phi_0$ = 275.26  ( +0.28, -0.30 )
%$\alpha$ = 19.55  ( +8.41, -8.47 )
%$\zeta$ = 27.47  ( +11.65, -11.85 )

\subsection{Timing solution}
\begin{figure*}
\centering
\begin{subfigure}{0.9\textwidth}
    \includegraphics[width=\linewidth]{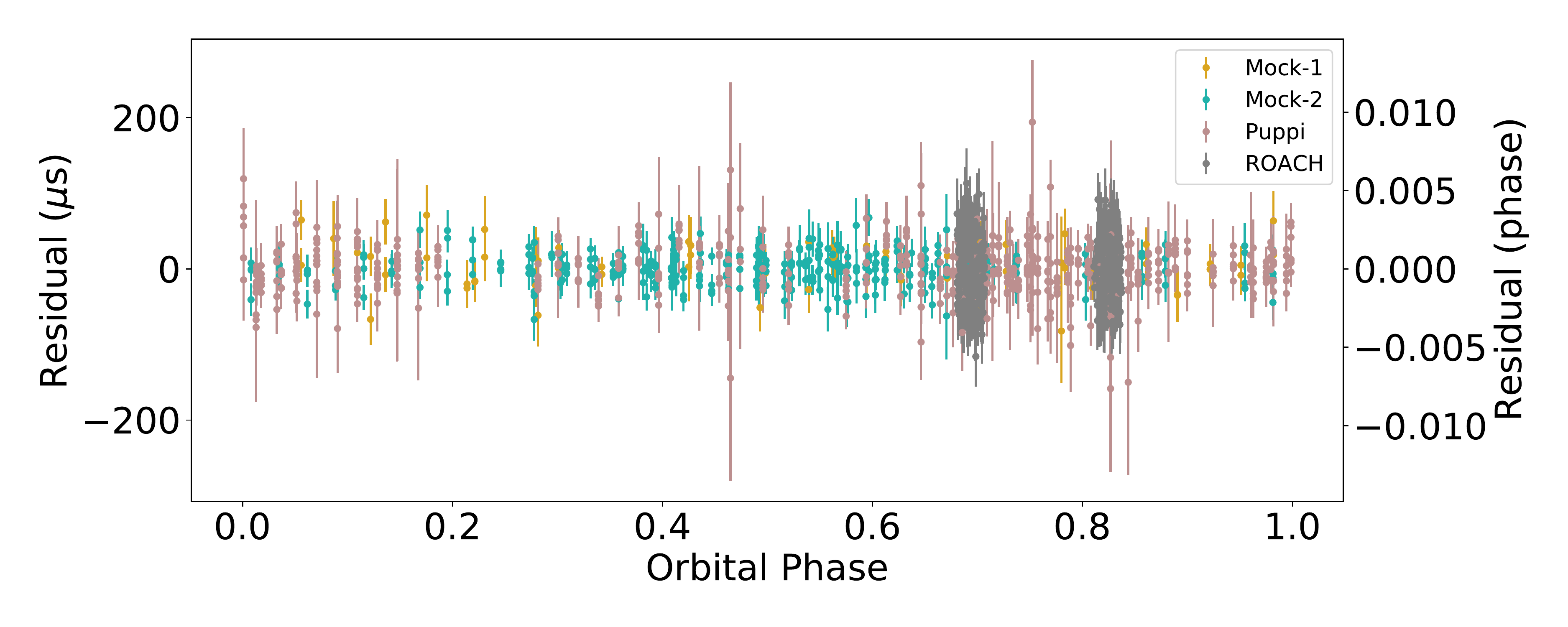}
\end{subfigure}
\hfill
\begin{subfigure}{0.9\textwidth}
    \includegraphics[width=\linewidth]{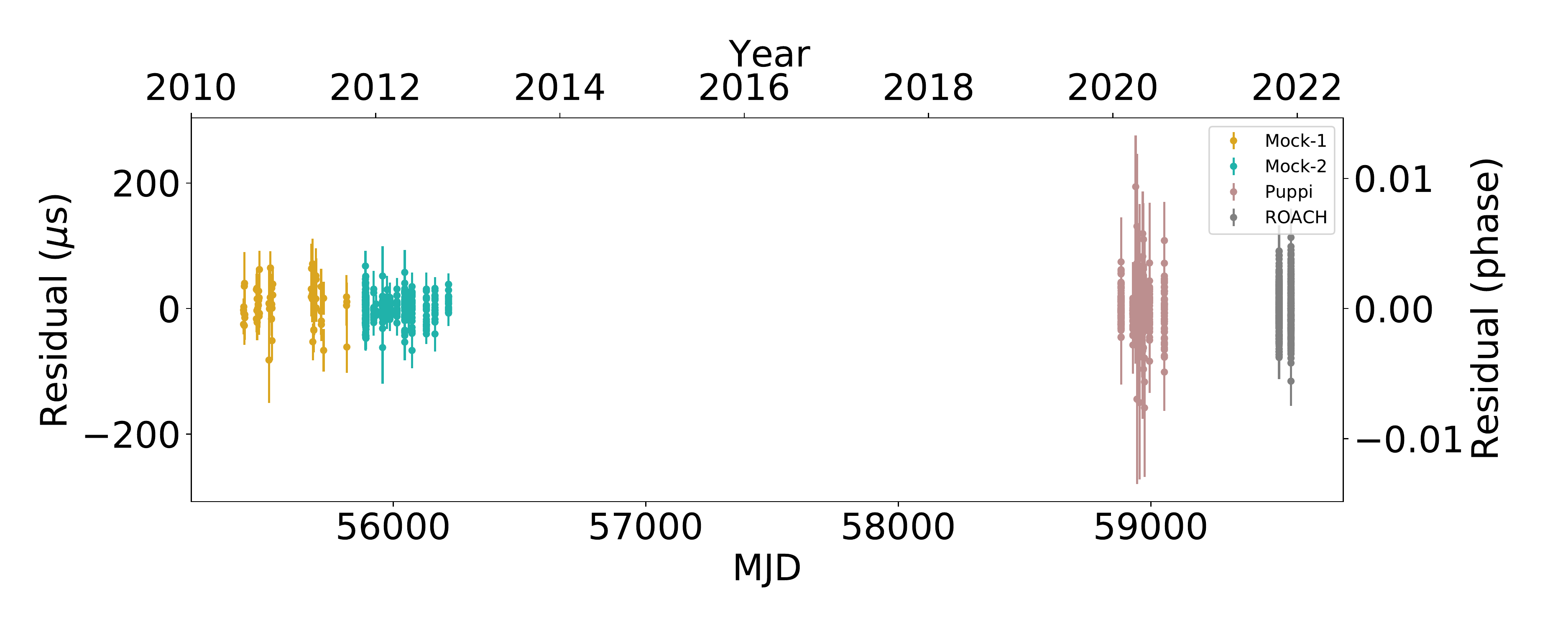}
\end{subfigure}
\caption{Post-fit residuals of ToAs from Arecibo data (with Mock and PUPPI backends) and FAST data (with ROACH backend) as a function of year (below) and orbital phase of pulsar (top). Different colours show residuals for different backends. Points labelled Mock-1 and Mock-2 show different sets of timing observations with the Mock spectrometers (see Table~\ref{table:observations}).}
\label{fig:residual-plots}
\end{figure*}

The timing solution of PSR~J1952+2630 resulting from the fit using the ELL1H+ and DDGR binary models is shown in Table~\ref{tab:timing_solution_J1952}. Compared to the previous work on this pulsar, we have extended the timing baseline by more than nine years and increased the number of ToAs by a factor of 3.5.
Fitting a total 1475 ToAs, we obtain a weighted RMS of $\sim$ 24.2 $\mu \rm s$ with a reduced $\chisquare$ of 1.01 (having 1457 degrees of freedom) using the DDGR model. The uncertainties of ToAs from different backends were individually scaled such that the reduced-$\chisquare$ of each data set is 1. Figure~\ref{fig:residual-plots} shows the post-fit residuals of ToAs as a function of the epoch of observation and the orbital phase of the binary.

\subsection{Astrometry}

Relative to the values published by \cite{2014MNRAS.437.1485L}, we achieved an order-of-magnitude improvement in Right Ascension (RA) and an improvement of a factor of 2 in Declination (DEC). We were also able to refine the proper motion of the system in RA by an order of magnitude and measure the proper motion in DEC for the first time. The total proper motion of the system is 5.89(23) mas yr$^{-1}$. The distance prediction from the DM of this pulsar is 9.56 kpc from the NE2001 electron density model \citep{2002astro.ph..7156C} and 10.03 kpc from the YMW16 model \citep{2017ApJ...835...29Y}. Considering the NE2001 value with a 20\% error, we constrained the transverse velocity of the system to 267(53) km s$^{-1}$.
%We also get an order of magnitude improvement in the mass function of the system, $f$ = 0.15317962(43). 

\begin{table*}
\caption{Timing solution of PSR~J1952+2630 with DDGR and ELL1H+ models. Values in brackets are 1$\sigma$ uncertainties, and values without uncertainties are derived from the corresponding binary model. Values in square brackets are not fit, but derived assuming GR. \footnotesize{$^a$ Assumed value derived from the $\rm sin\it i$ parameter in the DDGR solution. $^b$ Assuming the distance and uncertainty of NE2001.}}
\label{tab:timing_solution_J1952}
\centering
\footnotesize
\setlength{\tabcolsep}{15.0pt}
\renewcommand{\arraystretch}{1.1}
\begin{tabular}{l c c}

\hline\hline 
\multicolumn{3}{c} {PSR J1952+2630}                                                  \\
\hline 
Binary Model                                                            &   DDGR       &   ELL1H+                                                               \\
\hline
Observation and data reduction parameters \\
\hline
Reference epoch (MJD)                                                   &   55407.0    &   55407.0                                                     \\
Span of timing data (MJD)                                              &   55407.146 -  59554.385 &   55407.146 -  59554.385  \\
Solar system ephemeris                                                  &   DE436   &   DE436                                                             \\
Terrestrial time standard                                               &   UTC(NIST)     &   UTC(NIST)                                                       \\
Time units                                                              &   TDB    &   TDB                                                                 \\
Number of ToAs                                                          &   1475     &   1475                                                     \\
Weighted residual RMS ($\mu$s)                                                  &   24.201      &   24.198                                                               \\
$\chisquare$             & 1466.43 & 1466.10 \\
Reduced $\chisquare$         & 1.0065 & 1.0069 \\
\hline
Astrometric and spin parameters\\
\hline
Right Ascension, $\alpha$ (J2000)                                       &   19:52:36.840244(57)   & 19:52:36.8402341(58)                                                     \\
Declination, $\delta$ (J2000)                                           &   26:30:28.0840(11)  & 26:30:28.0843(11)                                                        \\
Proper motion in $\alpha$, $\mu_\alpha$ (mas yr$^{-1}$)                 &   $-$2.58(17)  &   $-$2.57(17)                                                          \\
Proper motion in $\delta$, $\mu_\delta$ (mas yr$^{-1}$)                 &   $-$5.29(24)  &   $-$5.38(25)                                                            \\
Spin frequency, $\nu$ (Hz) 
        & 48.2337753923167(47)   & 48.2337753923193(49)  \\
Spin frequency derivative, $\dot{\nu}$ ($10^{-15}$Hz s$^{-1}$) 
        &  $-$9.93790(9)  & $-$9.93791(9)  \\
Dispersion Measure, DM (pc cm$^{-3}$)                                   &   315.3168(17)     &   315.3166(17)                                                          \\
\hline
Binary parameters  \\
\hline
Orbital period, $\Pb$ (days)                                            &   0.39188050(21)                                                         &   0.391878632876(93)      \\
Projected semi-major axis, $x_{\rm p}$ (lt-s)                           &   2.7981770(18)     &   2.7981793(49)                                                           \\
Orbital eccentricity, $e$                                               &   4.226(99)$\times 10^{-5}$                                          &    ...       \\
Longitude of periastron, $\omega$ (deg)                                 &   289.56(1.64)                                                                &  ...    \\
Epoch of periastron, $T_0$ (MJD)                             &   55407.6178(18)                                                       &   ...        \\
Epoch of ascending node, $T_{\rm asc}$ (MJD)    
         &   ...   &   55407.3025831(1) \\
$\epsilon_1$  & ... & $-$0.000039(2) \\
$\epsilon_2$ &   ...  &   0.000013(1) \\
Advance of periastron, $\dot{\omega}$ (deg/yr)                     &   [1.5998941]                                                        &   1.80(26)        \\
Shapiro shape, $s = \sin i$                                                                  &    [0.9495500]     &    ...  \\
Orbital period derivative, $\dot{P}_{\rm b}$ (10$^{-12}$ s s$^{-1}$)    &   [$-$0.0943648]                                             &   $-$0.096(46)         \\
Orthometric ratio, $\varsigma$                                                 &  ...                &    0.722852$^a$                           \\
Orthometric amplitude, $h_3$, ($\mu$s)                                     &  ...        &   1.49(61)    \\ 

Total binary mass, $M_{\rm T}$ ($\rm M_\odot$)                                             &   2.19(38)    & [2.61$^{+0.58}_{-0.54}$]                                                             \\
Companion mass, $M_{\rm c}$ ($\rm M_\odot$)                                 &   0.95(12)    
  &   ... \\
\hline
Derived parameters\\
\hline
Galactic longitude, $l$ ($\deg$)  & 63.254   & 63.254\\
Galactic latitude, $b$ ($\deg$) &  $-$0.376 &  $-$0.376 \\
DM derived distance from NE2001 model (kpc) & 9.5647 & 9.5647\\
DM derived distance from YMW16 model (kpc) & 10.03 & 10.03 \\
Galactic height, z (kpc)    & $-$0.063(12) & $-$0.063(12) \\
Magnitude of total proper motion, $\mu$ (mas yr$^{-1}$)   &   5.89(23) & 5.96(24) \\
Transverse velocity$^b$, $v_{\rm T}$  (km s$^{-1}$) & 267(53) & 270(54) \\
%20% uncertainty of DM distance used..
Spin period, $P$ (s)                                                    &   0.020732360091375(2)        &   0.020732360091373(2)                                           \\
1st Spin period derivative, $\dot{P}$ ($10^{-18}$ s s$^{-1}$)                      &   4.27162(4)                  &   4.27162(4)                        \\
Intrinsic spin period derivative, $\dot{P}$ ($10^{-18}$)& 4.270(3) & 4.270(3)  \\
Characteristic age, $\tau_c$ (Myr)     & 76.91 &   76.92 \\
Surface magnetic field, $B_{\rm s}$ ($10^9$G)   &  9.49 & 9.49   \\
Spin-down energy, $\dot{E}$ ($10^{33}$erg s$^{-1}$)     & 18.92 & 18.92\\

Kinematic contribution to $\dot{P}_{\rm b}$ ($10^{-15}$ s s$^{-1}$) & 1.8$^{+5.1}_{-3.4}$ & 2.5$^{+5.2}_{-3.5}$ \\

Mass function, $f(M_{\rm p})$ ($\rm {M}_\odot$)           &    0.15317962(43)                                                              &   0.15318146(81)  \\
Pulsar mass, $M_{\rm p}$ ($\rm M_\odot$)                                 &   1.24(26)                  &    ...            \\
Einstein delay, $\gamma$ (s)                                            &   2$\times 10^{-7}$           & ...                                    \\
Relativistic orbital deformation, $\delta_{\theta}$ (10$^{-6}$)         &   4.674351                                                & ...                 \\
Relativistic orbital deformation, $\delta_{\rm r}$ (10$^{-6}$)          &   4.4226831                                                    & ...             \\ 
\hline
\hline
\end{tabular}
\clearpage
\end{table*}

\subsection{Kinematic effects on the spin and orbital periods}
\label{sec:kinematic_effects}
The observed arrival times of the pulses are affected by the relative accelerations between the pulsar binary and the Solar System. These include a) the Shklovskii effect \citep{1970SvA....13..562S}, an apparent centrifugal acceleration, and b) the line-of-sight acceleration acting on the pulsar due to the difference in the Galactic acceleration on the pulsar and the Solar System.

The  Shklovskii effect, $a_{\rm shk}$ is simply $\mu^2d$. Here, $\mu$ is the total proper motion and $d$ is the distance to the pulsar. 

The galactic acceleration includes contributions from both the Galactic rotation and vertical accelerations to the disk of the galaxy ($a_{\rm gal} = a_{\rm gal,disc} + a_{\rm gal,rot}$). These are given by \cite{1991ApJ...366..501D}, \cite{1995ApJ...441..429N}, and \cite{2009MNRAS.400..805L} as follows:

\begin{eqnarray}
     a_{\rm gal, rot} & = & -\frac{\Theta_0^2}{R_0 c}\left(\rm cos \emph{l}+\frac{\beta}{\beta^2+\rm sin^2\emph{l}}\right)\rm cos \, \it{b} ,\\
     a_{\rm gal, disc} & = & -\left(2.27 z_{\rm kpc} + 3.68 (1-e^{-4.3z_{\rm kpc}})\right) |\rm sin\, \it{b}| ,
\end{eqnarray}
where $\beta \equiv \left({d/R_0}\right)\rm cos \emph{b}- \rm cos\emph{l}$ and $z_{\rm kpc} \equiv |d \rm sin\emph{b}|$ in kpc. $R_0$ is the distance to the galactic centre, taken as 8.275(34) kpc, and $\Theta_0$ is the galactic rotation velocity that is taken to be 240.5(41) km s$^{-1}$ \citep{2021A&A...647A..59G,2021A&A...654A..16G}. 
PSR~J1952+2630 is close to the Galactic disk ($b = -0.376 \deg$); this means that $a_{\rm gal, rot}$ will be the dominant component by far, and the vertical acceleration $a_{\rm gal, disc}$ is almost perpendicular to the line of sight.

The observed spin period derivative of the pulsar ($\Pdotobs$) includes these acceleration contributions as additional Doppler shift effect,
\begin{equation}
    \frac{\Pdotobs}{P} = \frac{\Pdotint}{P} + \frac{a_{\rm shk} + a_{\rm gal}}{c},
\end{equation}
 where $\Pdotint$ is the intrinsic spin period derivative of the pulsar. Considering the astrometric parameters from the fit with DDGR model, we estimated the cumulative kinematic contribution in the spin period derivative, $P\left(\frac{a_{\rm shk} + a_{\rm gal}}{c}\right)$ as $1.14\times 10^{-21}$ s s$^{-1}$. This is much smaller than $\Pdotobs$. When we subtract it from the latter value, we therefore obtain $\Pdotint$ of the system of $4.2704(32) %^{+0.0021}_{-0.0032}
 \times 10^{-18}$ s s$^{-1}$, which is very similar to the uncorrected value. With this value, we estimated the surface magnetic field, $B_{\rm s}$ of 9.49 $\times 10^9$ G, characteristic age, $\tau_c$ of 76.91 Myr, and the rate of loss of rotational energy $\dot{E}$ of 18.92 $\times 10^{33}$erg s$^{-1}$ for this pulsar (all estimated using the relations in \citealt{2012hpa..book.....L}). The detailed study of the spin evolution of this system presented by \cite{2014MNRAS.437.1485L} therefore remains valid.
 
Similar accelerations from proper motion (Shklovskii effect), galactic disk, and galactic rotation also contribute in the observed orbital period derivative measurement,
\begin{equation}
    \Pbdotobs = \Pbdotint + \dot{P}_{b,\,\rm shk} + \Pbdotgal .
\end{equation}

\begin{figure}
\centering
        \includegraphics[width=\columnwidth]{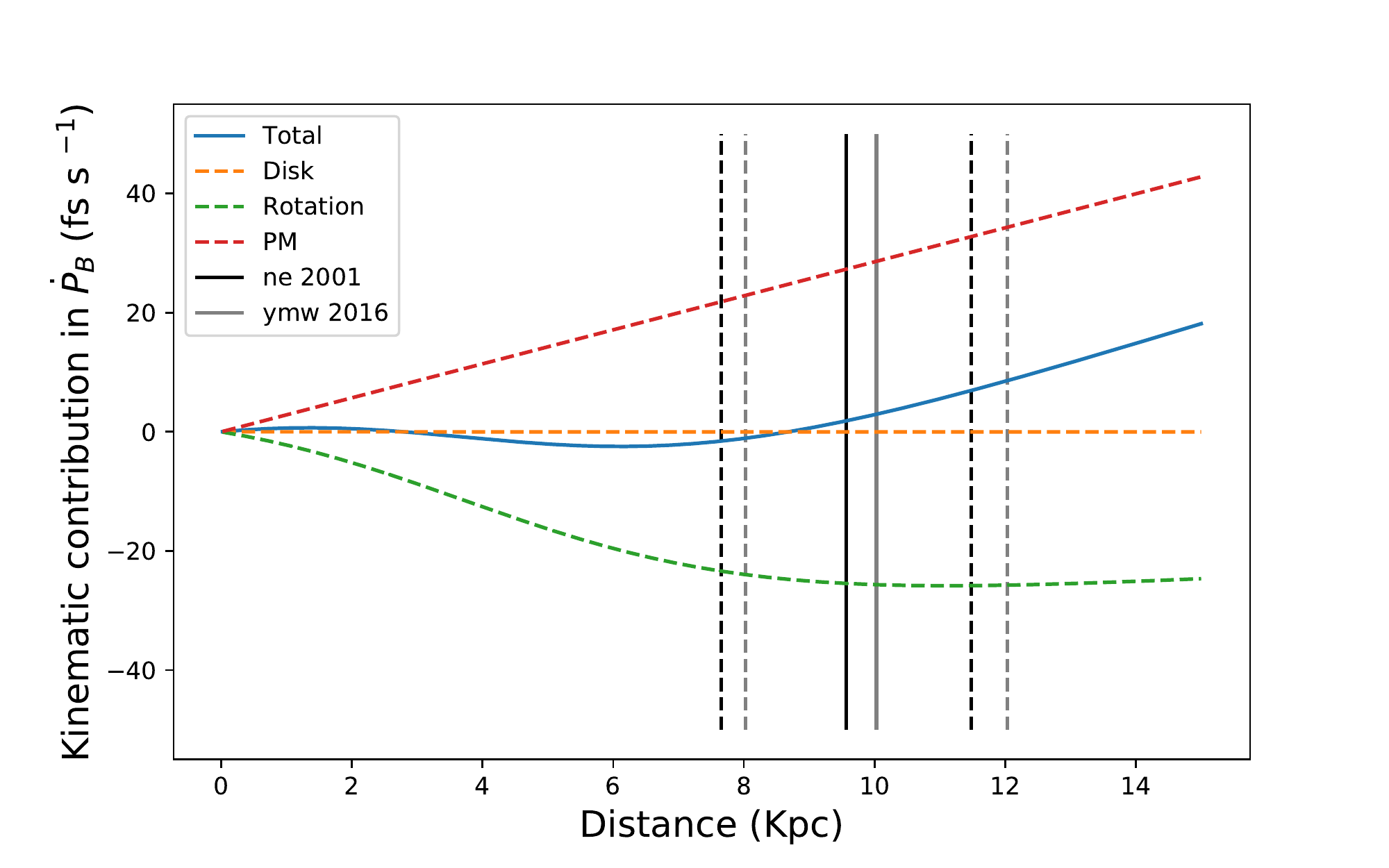}
        \caption{Total kinematic contribution to the observed orbital decay, $\dot{P}_{b, \rm gal}$  , shown as a function of distance to the pulsar in the blue curve. The red, orange, and green curves represent individual contributions due to galactic disk acceleration, galactic rotation (or differential) acceleration, and proper motion of the pulsar, respectively, with errors in $R_0$ and $\theta_0$ providing negligible contributions. The vertical black line marks the DM distance from the NE2001 model. The 20$\%$ uncertainty in the model is shown as dashed black lines. Grey lines show an estimate of the DM distance from the YMW16 model. The maximum uncertainty in total kinematic contribution is $5.1 \rm fs \, \rm s^{-1}$.}
        \label{fig:pbdot-dist}
\end{figure}

Figure~\ref{fig:pbdot-dist} shows these contributions as a function of the distance to the pulsar. When we assume a 20$\%$ error in the distance prediction of the NE2001 model, we obtain a maximum uncertainty of 1.91 kpc in its measurement. This uncertainty in distance implies an uncertainty in the estimation of these kinematic contributions in $\dot{P}_{b}$, 1.8$^{+5.1}_{-3.4} \rm \, \rm fs \, \rm s^{-1}$ (see the dashed lines in figure~\ref{fig:pbdot-dist}). Because the curves are relatively flat, any improvement in the distance measurement in future will not change this constraint significantly. As we discuss in detail in Section~\ref{sec:lab}, this contribution also presents the final constraint in the $\Pbdotobs$ precision in future. 

\subsection{Advance of periastron}
Using the ELL1H+ binary model, we obtained a rate of the  change in the longitude of periastron of the binary ($\dot{\omega}$) of 1.80(26) deg yr$^{-1}$. When GR is assumed as a theory of gravity, this PK parameter depends on the Keplerian parameters and the total mass of the system ($M_{\rm T}$) as follows:
\begin{equation}
\label{eq:omdot}
    \dot{\omega} = 3{T_\odot}^{{2}/{3}}\left(\frac{\Pb}{2\pi}\right)^{{-5}/{3}} \frac{1}{1-e^2}   M_{\rm T}^{2/3},
\end{equation}
where masses are expressed in solar units. From this relation, we obtain $M_{\rm T} = 2.61^{+0.58}_{-0.54}$ \msun. This value is consistent with DDGR model fit for $M_{\rm T}$ of 2.19 (38) \msun.

\subsection{Shapiro delay}

In our ELL1H+ timing solution, we were also able to fit for the Shapiro delay.
When we assume GR, the orthometric parameters relate to $\Mc$ and $i$ as follows:
\begin{eqnarray}
    \varsigma & = & \frac{\sin i}{1 + \sqrt{1 - \sin^2i}}, \\
    h_3 & = & T_\odot \Mc \varsigma^3. \label{eq:h3}
\end{eqnarray}

Due to a weak Shapiro-delay signature in the data, we fixed the $\varsigma$ parameter to a constant value, 0.722, which corresponds to the best value of $\sin i$ from the DDGR solution (0.95), and only fit for $h_3$ parameter. From this, we obtain a 2$\sigma$ detection of $h_3$ parameter, $1.49(61)\, \mu$s. 

\subsection{Orbital period derivative}
\label{sec:pk-1}
The third PK parameter measured in this system is the orbital period derivative, $\Pbdot$. The ELL1H+ binary model fit gives a $\Pbdotobs$ of $-96(46)$ fs s$^{-1}$. 
Removing the kinematic contributions (discussed in Section~\ref{sec:kinematic_effects}) from $\Pbdotobs$, we obtain $\Pbdotint =  -97^{+49}_{-51}$ fs s$^{-1}$.

The DDGR model does not directly fit for the orbital decay and instead fits for $\Mc$ and $M_{\rm T}$ using all relativistic effects in a consistent way. From the best-fit masses, it assumes that the orbital decay is due to quadrupolar GW emission predicted by GR, $\dot{P}_{\rm b,GR}$ , as follows: 
\begin{equation}
\label{eq:pbdot}
    \dot{P}_{\rm b,GR} = - \frac{192 \pi}{5} T_\odot^{5/3}\left(\frac{\Pb}{2\pi}\right)^{-5/3}f(e)\frac{\Mp \Mc}{(\Mp+\Mc)^{1/3}} ,
\end{equation}
where
\begin{equation}
    f(e) = \frac{1+\frac{73}{24}e^2 + \frac{37}{96}e^4}{(1-e^2)^{7/2}}.
\end{equation}
Using this model, the derived $\dot{P}_{\rm b,GR}$ is $-94(27)$ fs s$^{-1}$. 

\section{Mass measurements}
\label{sec:measurements}

With the DDGR binary model, we directly obtain $M_{\rm T}$ of 2.19(38) \msun \, and $M_{\rm c}$ of 0.95(12) \msun. This gives $M_{\rm p}$ of 1.24(26) \msun. Using these values along with the mass function of this system, we obtain an orbital inclination of 71.72 deg.

\begin{figure}
\centering
\begin{subfigure}{0.45\textwidth}
    \includegraphics[width=\linewidth]{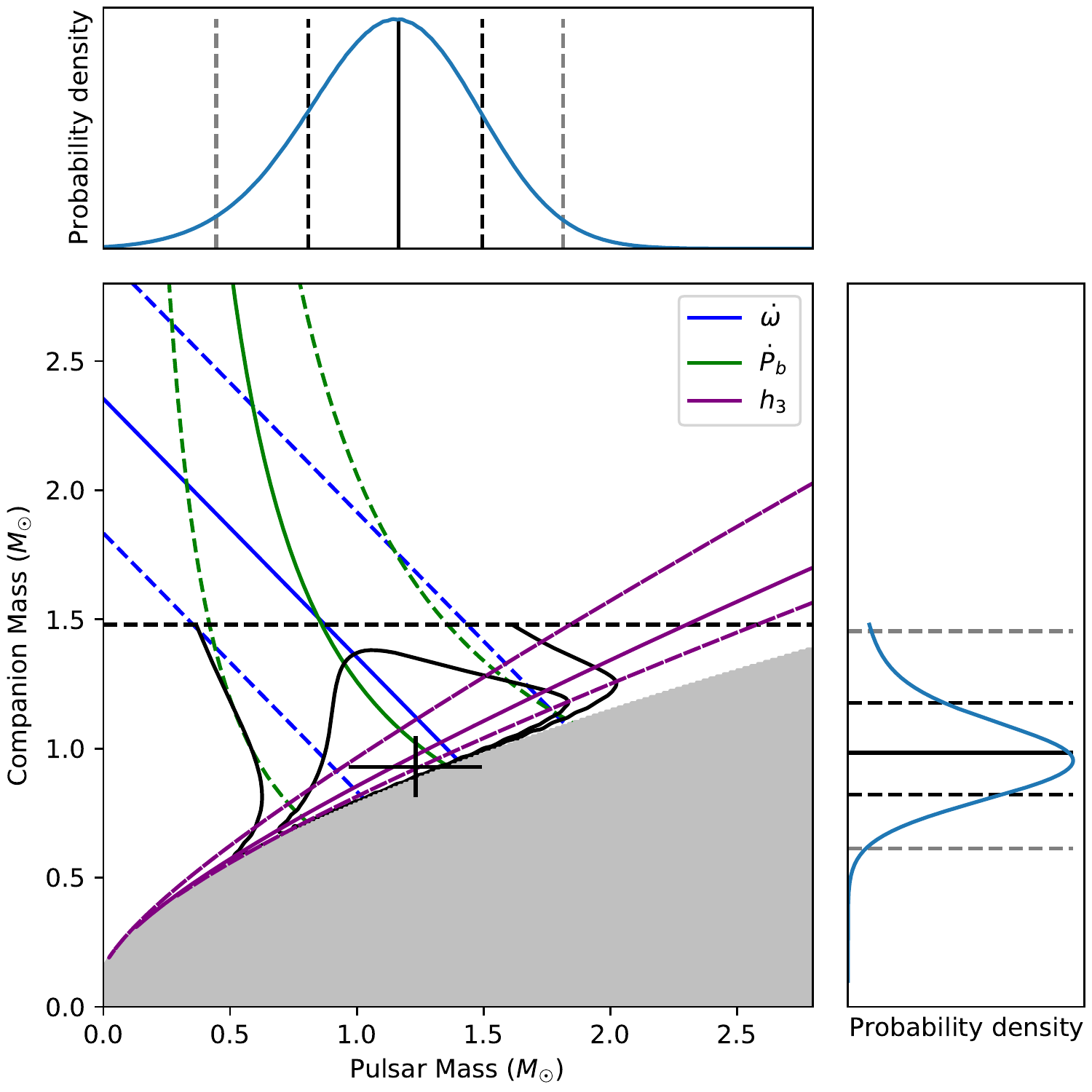}
\end{subfigure}
\hfill
\begin{subfigure}{0.45\textwidth}
    \includegraphics[width=\textwidth]{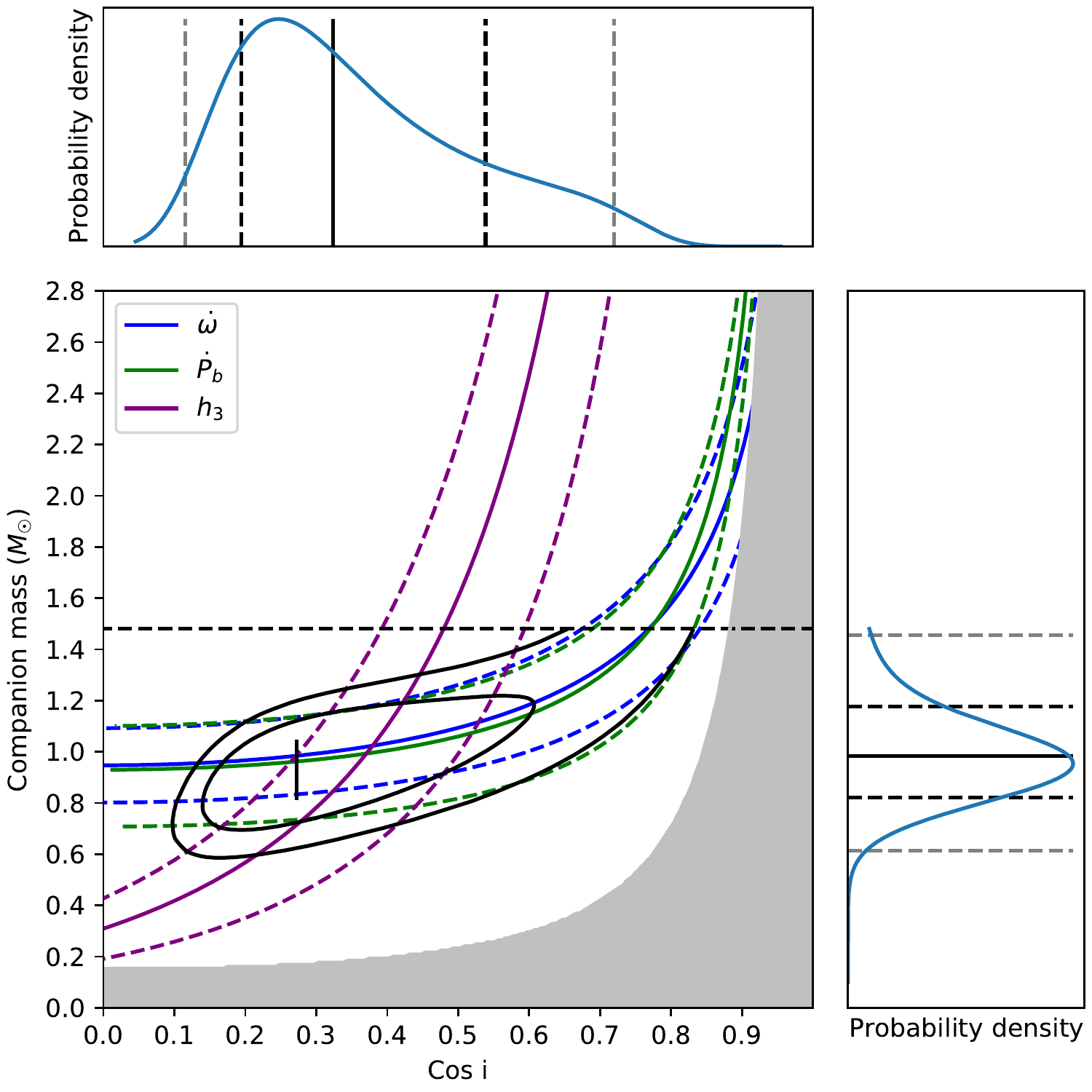}
\end{subfigure}
\caption{Constraints on the $\Mc$, $M_{\rm p}$, and $\cos i$ derived from the measurement of three relativistic PK parameters are shown. The dashed black line shows the maximum $\Mc$ constraint of 1.48 \msun. Coloured dashed lines depict the 1$\sigma$ uncertainty on the measurements. Black contours show the likelihood or 2D PDFs at 68 $\%$ (inner) and 95 $\%$ (outer) confidence from the $\chisquare$ maps on $\cosi-\Mc$ and $\Mp-\Mc$ grids, derived from the DDGR fit. The corner plots represent the marginalised 1D PDFs of each of these measurements, and vertical lines in them mark the median, 1$\sigma$, and 2$\sigma$ estimates. The black crosses indicate the best masses from the DDGR solution and the best inclination derived from the mass function using these masses.}
\label{fig:mp-mc-cosi-planes}
\end{figure}

We now determine whether the three PK measurements from ELL1H+ solution are, according to GR, consistent with these two masses and with each other. The mass constraints that result from $\dot{\omega}$, $\Pbdot$, and $h_3$ and their 1$\sigma$ uncertainties can be represented on the pulsar and companion mass plane ($\Mp$-$\Mc$) and on the companion mass and inclination planes ($\cos i$ - $\Mc$); see Figure~\ref{fig:mp-mc-cosi-planes}. Because the curve of $h_3$ parameter cross $\dot{\omega}$ and $\Pbdot$ curves relatively steeply in these plots, this measurement of Shapiro delay gives us the stronger constraints on the individual masses and inclination. All PK parameters agree in a particular region of the maps, which coincides with the masses derived from the DDGR solution (given by the black crosses). This means that GR provides a satisfactory and self-consistent description of all the relativistic effects in this system. However, given the low precision of the measurement of all PK parameters, this test of GR is not particularly constraining.

 In order to better constrain the masses and avoid any correlations that might be present between the parameters, we created a self-consistent $\chisquare$ map on a 2D grid of $\cos i$ and $\Mc$. For each point in the grid, we estimated $M_{\rm T}$ using the mass function, and kept it and $\Mc$ fixed in a DDGR solution, using \TEMPO to fit for all other parameters. The resulting $\chisquare$ values for each point on the $\cos i$-$M_{\rm c}$ grid were then recorded and converted into a 2D probability distribution function (PDF), following \cite{2002ApJ...581..509S},
\begin{equation}
p(\cos i, M_{\rm c}) \propto e^{\frac{\chisquare_{\rm min} - \chisquare}{2}},
\end{equation}
where $\chisquare_{\rm min}$ is the minimum $\chisquare$ value in the whole grid. A corresponding 2D PDF grid for the $\Mp-\Mc$ plane was also derived using the mass function. Importantly, we did not sample values of $\Mc$ above $1.5 \rm M_{\odot}$ because that is the upper mass limit for a rigidly rotating WD.

The solid black lines in Figure~\ref{fig:mp-mc-cosi-planes} show contours that include 68\% and 95\% of the total probability in these 2D PDFs. To determine the resulting constraints for $M_{\rm p}$,  $M_{\rm c}$, and $\cos i$, these 2D PDFs were marginalised and converted into 1D PDFs. Following this, we obtain the  median value and 1$\sigma$ uncertainty on $M_{\rm p}$ of 1.20$^{+0.28}_{-0.29} \rm M_{\odot}$, $\Mc$ of 0.97$^{+0.16}_{-0.13} \rm M_{\odot}$, and inclination of 70.55$^{+6.15}_{-11.02} \deg$; these are, again, consistent with the solution from the DDGR model.

Using the DDGR model, we can fit for contributions to the variation of the orbital period other
than the GR orbital decay, using the XPBDOT parameter, which quantifies $\Pbdotx =\Pbdotobs - \Pbdotgr$. For this parameter, we obtain a null result,
implying that the kinematic contributions are certainly smaller than our current measurement uncertainties: The predicted kinematic contribution in $\dot{P}_\mathrm{b}$ of 1.8$^{+5.1}_{-3.4} \rm \, \rm fs \, \rm s^{-1}$ is indeed an order of magnitude smaller than the measured uncertainty in $\Pbdot$ of 46 $\rm fs \, \rm s^{-1}$ (from the ELL1H+ model).

\begin{figure}
\centering
        \includegraphics[width=\columnwidth]{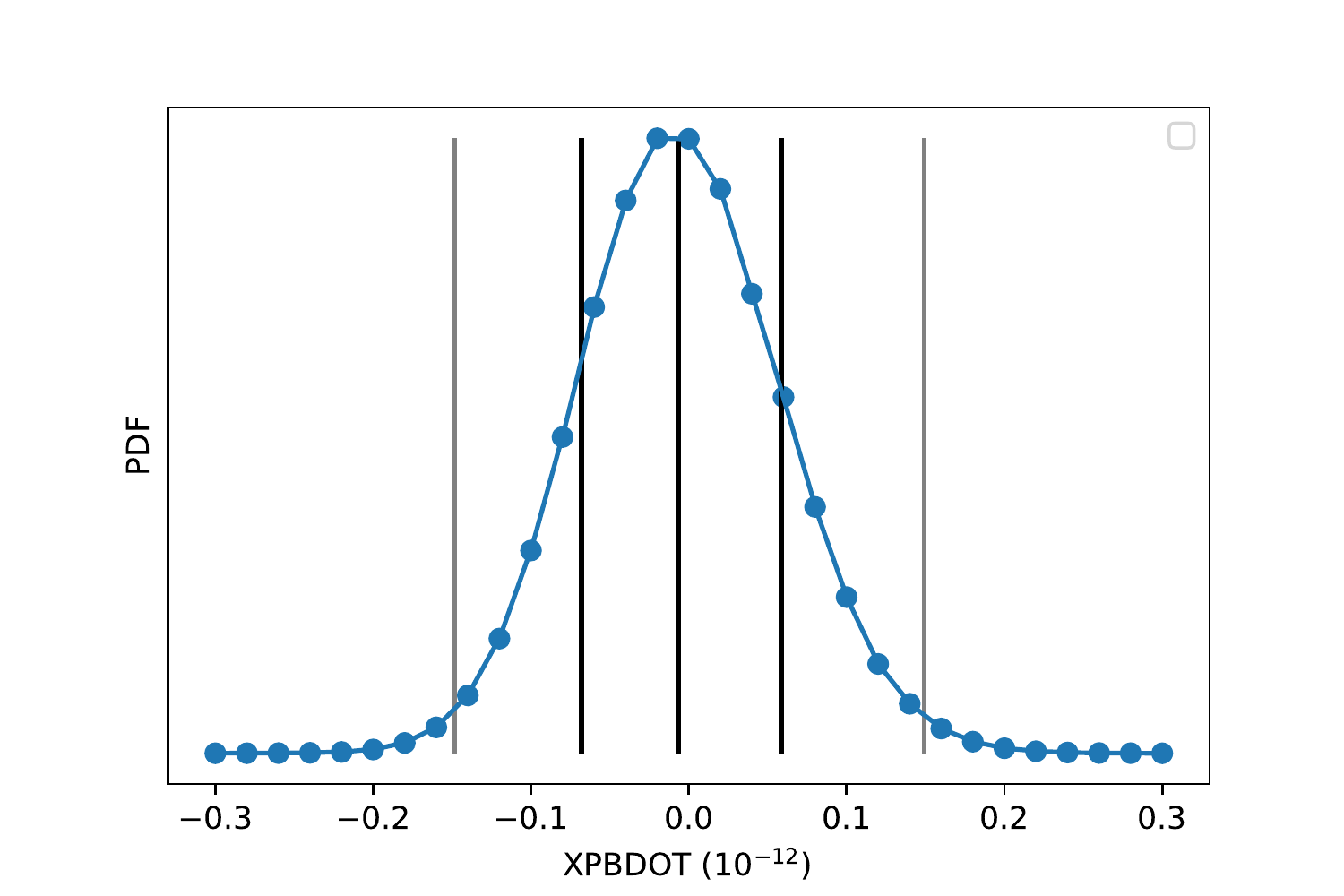}
        \caption{1D probability density function of $\Pbdotx$ derived from the best-fit $\chisquare$ of the DDGR solution. Blue dots represent the grid points over which the \TEMPO \, fit was made, and the blue curve shows the interpolated function. The vertical black lines represent the median and 1$\sigma$ values, and grey lines show the 2$\sigma$ constraint. We obtain a 1$\sigma$ estimate of $\Pbdotx$ as $-6^{+58}_{-68}\times \rm fs\, \rm s^{-1}$.}
        \label{fig:xpbdot-sim}
\end{figure}

In order to estimate $\Pbdotx$ precisely, we mapped a 3D grid of $\Mc$, $\cosi$, and $\Pbdotx$.
We determined the best-fit $\chisquare$ over each point in this grid using the DDGR binary model and then estimated the marginalised probability distribution for $\Pbdotx$ (as shown in Figure~\ref{fig:xpbdot-sim}). 
We obtain a median value with a 1$\sigma$ uncertainty of $-6^{+58}_{-68} \rm fs\, \rm s^{-1}$ on $\Pbdotx$, implying a negligible value symmetric around zero. In the future, when more precise mass measurements are available, this method will be useful for robust estimates of $\Pbdotx$ and its uncertainty. As we show below, this is important to constrain alternative gravity theories.

\section{PSR~J1952+2630 as a gravitational laboratory}
\label{sec:lab}
In addition to quadrupole GW emission implied by GR, other alternative theories of gravity can also have observable implications. One common signature predicted by several of these theories but not by GR is the emission of dipolar GWs \citep{1975ApJ...196L..59E,1996PhRvD..54.1474D}. This emission, if present, will be observable as an additional contribution to the orbital decay. 

Thus, for any measurement of orbital decay, we can subtract the contributions from kinematic effects and the quadrupole GW emission to estimate the remaining contribution (also known as the excess orbital decay, $\Pbdotxs$), which is an upper limit on the dipolar GW contribution.

From the current timing of PSR~J1952+2630, we estimated this quantity by subtracting the predicted kinematic contributions from the $\Pbdotx$ estimate (calculated from the 3D $\chisquare$ map discussed in Section~\ref{sec:measurements}), which does not include the quadrupolar GW contribution, as follows:
\begin{equation}
    \Pbdotxs = \Pbdotx - \Pbdotgal - \Pbdotshk .
\end{equation}
We obtain $\Pbdotxs$ of $ 4.2^{+70.2}_{-73.1} \rm fs \, \rm s^{-1}$.
This value is negligible, which means that no dipolar GW emission is detectable and that GR passes the test posed by the measured PK parameters for PSR~J1952+2630.

Scalar-tensor theories (STT) of gravity predict the emission of scalar waves in asymmetric systems that have significantly different gravitational self-energies, such as NS-WD binaries.
Because of this, the scalar field in the theory will couple differently with both objects in the binary, which acquire a scalar charge (or effective coupling strength) given by $\alphap$ and $\alphac$. This is important because the dipolar GW emission depends on the difference between $\alphap$ and $\alphac$.

We can use the above prediction of $\Pbdotxs$ to place a constraint on the difference in effective scalar couplings of the pulsar and its companion, as predicted by STTs. Following equation 4 of \cite{2017ApJ...844..128C}, assuming a negligible eccentricity in this system, it can be calculated as follows:
\begin{equation}
    |\alphap-\alphac|^{2} < \delta\Pbdot^D \, \left(\frac{\Pb}{4\pi^2}\right)\left(\frac{1}{T_\odot \Mc}\right) \frac{\left(q+1\right)}{q},
\end{equation}
where q(=$\Mp/\Mc$) is the mass ratio and $\delta\Pbdot^D$ is the orbital decay from dipolar GW emission. Considering the 1$\sigma$ uncertainty in $\Pbdotxs$ as $\delta\Pbdot^D$, we obtain a limit on $|\alphap-\alphac| < 4.8 \times 10^{-3}$. This is not competitive with current tight constraints for other systems, estimated similarly from 1$\sigma$ error in $\Pbdotxs$: for PSR~J1738+0333 $|\alphap-\alphac| < 1.90 \times 10^{-3}$ \citep{2012MNRAS.423.3328F} and for PSR~J2222$-$0137 $|\alphap-\alphac| < 3.3 \times 10^{-3}$ \citep{2021A&A...654A..16G}. However, as we show below, the situation will improve significantly with continued timing.
%TG: Negligible errors from masses, not including them here.
%4.8(3)
%1.90(2)
\subsection{Simulations}

To determine the potential of this system in measuring precise component masses and testing GR, we simulated continued timing up to the year 2039. We first assumed a conservative scheme of orbital campaigns, once every two years until 2027 and then once every three years 
until 2039 with the FAST radio telescope. Our simulations are based on the timing precision of our current FAST data, which are about twice as precise as our Arecibo data. We therefore considered an RMS of 11 $\rm \mu \rm s$ for our simulations. At 48 ToAs every hour, every 9-hour orbit results in 432 simulated ToAs. Each of these orbital campaigns covers the short orbit fully, thus always inclduing the important orbital phases near superior conjunction, where the Shapiro delay is more pronounced. The masses and orbital parameters assumed in the simulation correspond to the best current DDGR fit.

\begin{figure}
\centering
    \begin{subfigure}{0.5\textwidth}
        \centering
    \includegraphics[width=0.8\columnwidth]{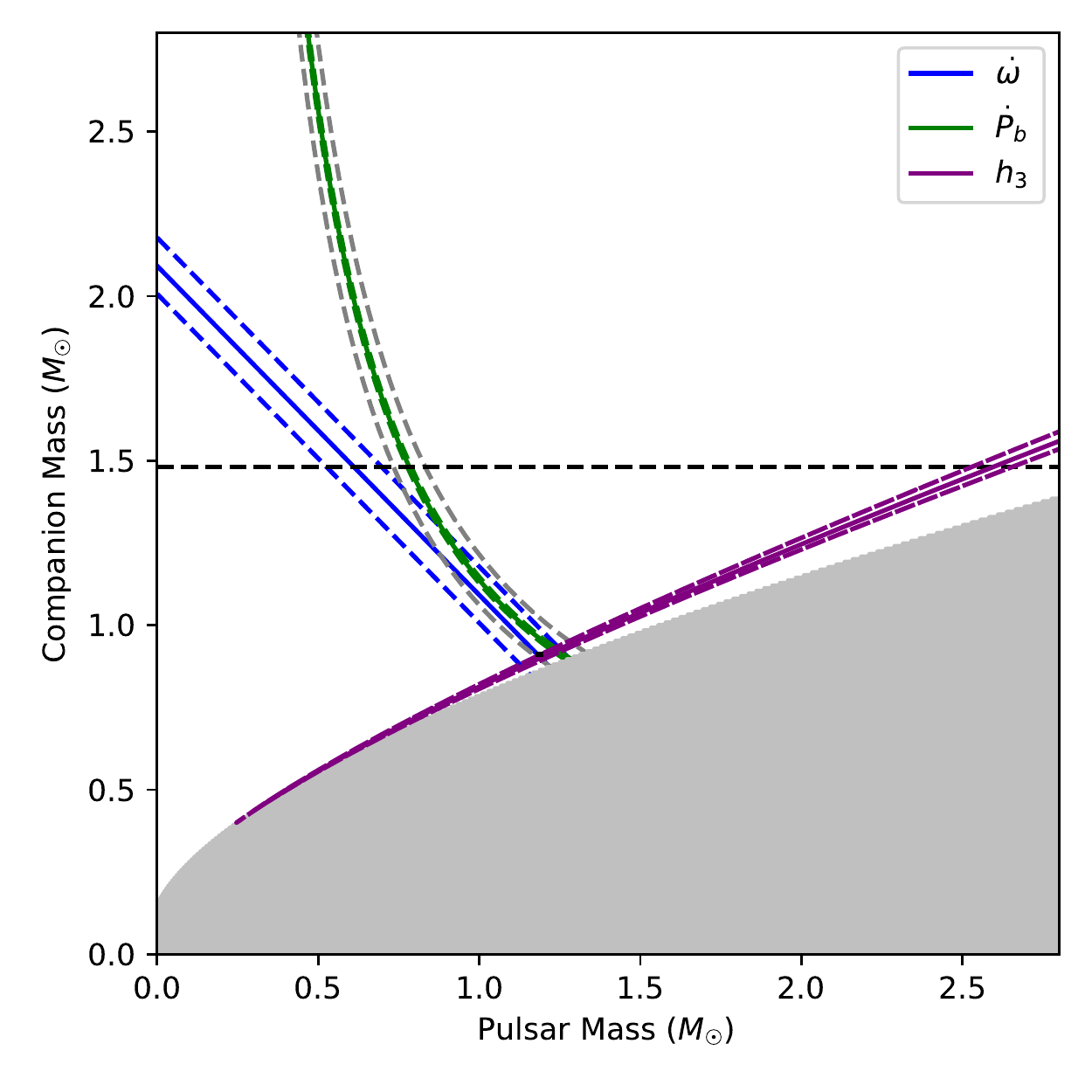}%
    \llap{\raisebox{0.85cm}{%  move next graphics to top right corner
    \hspace{-0.1ex}
      \includegraphics[height=3cm]{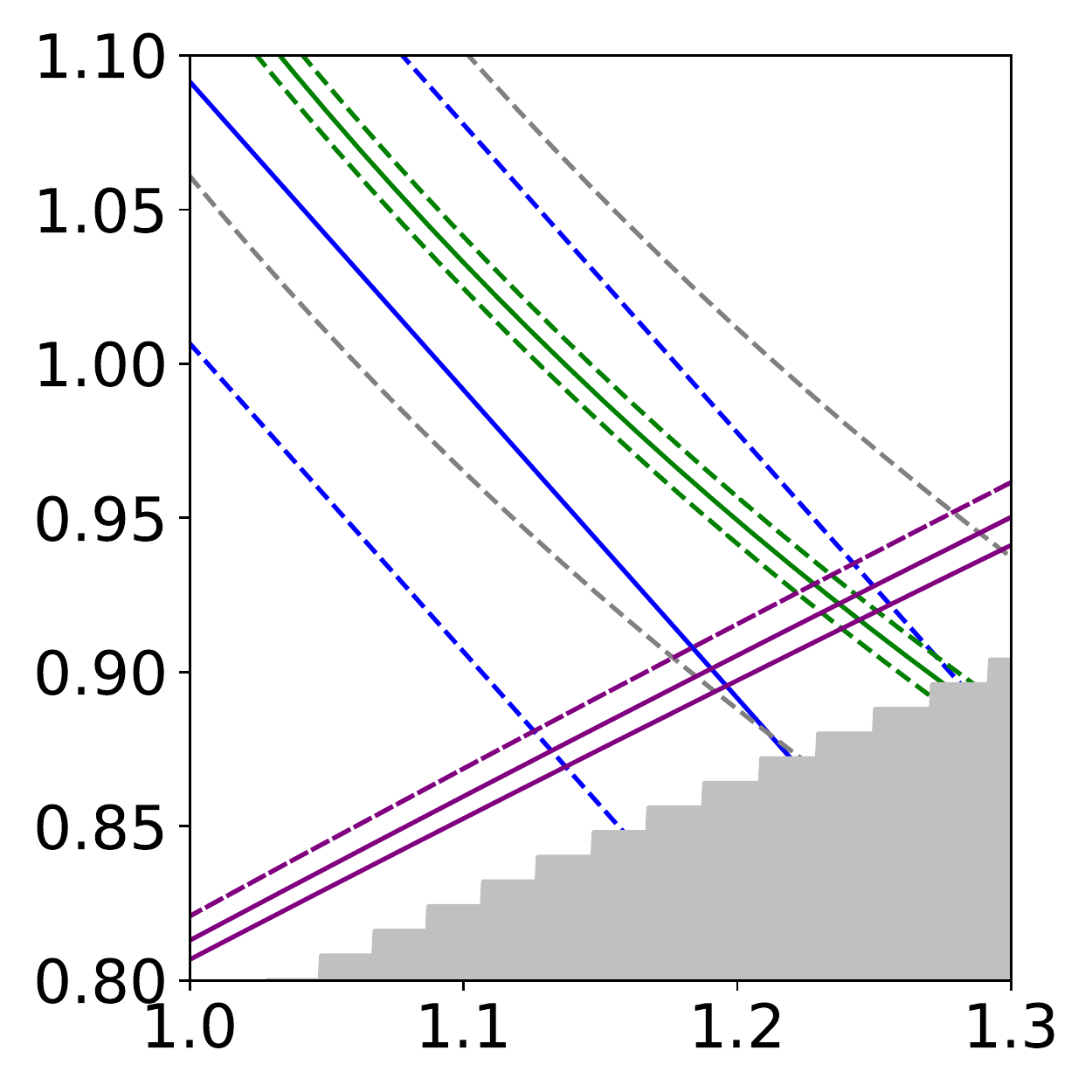}%
    }}
\end{subfigure}
\hfill
\begin{subfigure}{0.5\textwidth}
    \centering
    \includegraphics[width=0.8\columnwidth]{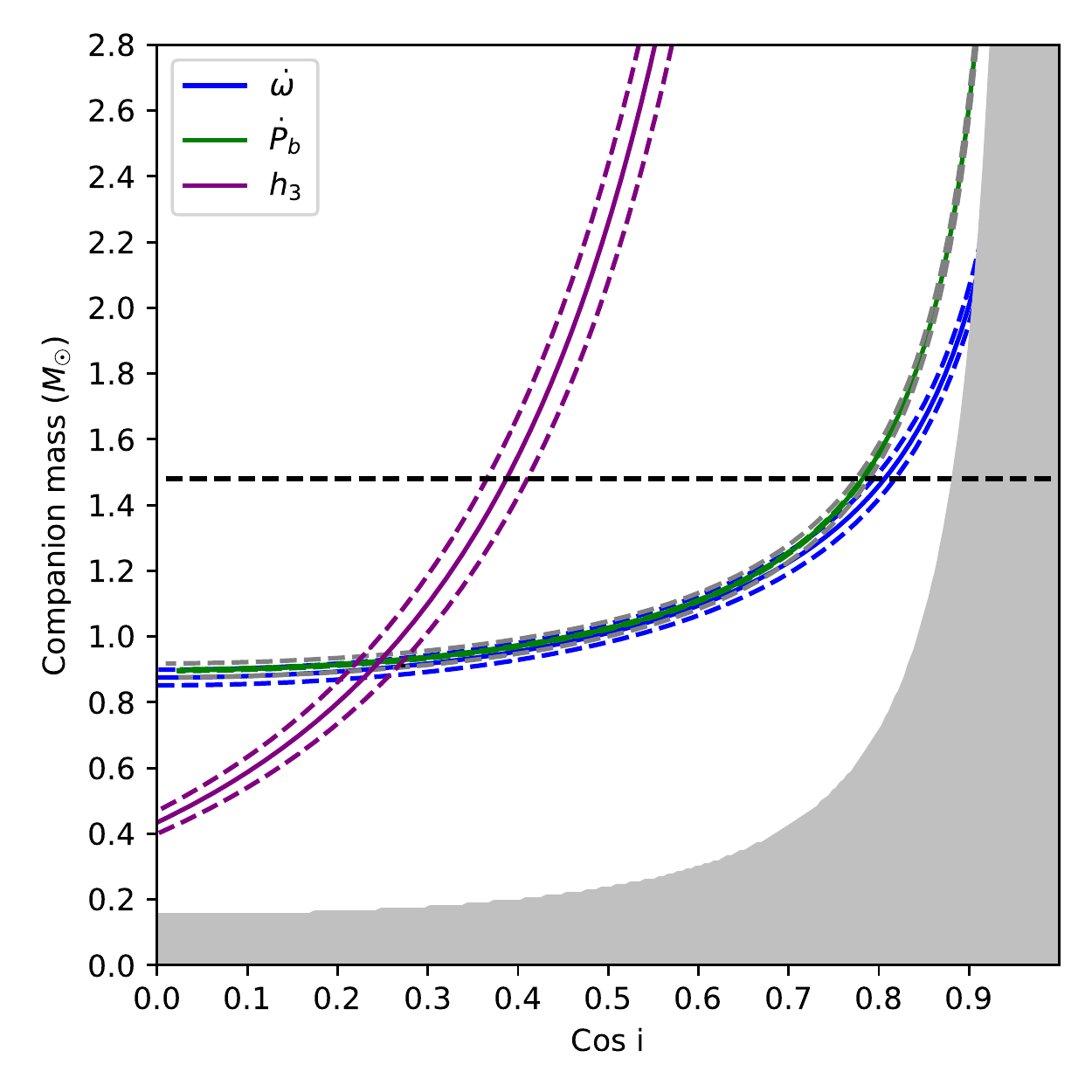}%
    \llap{\raisebox{0.8cm}{%  move next graphics to top right corner
    \hspace{-0.1ex}
      \includegraphics[height=3cm]{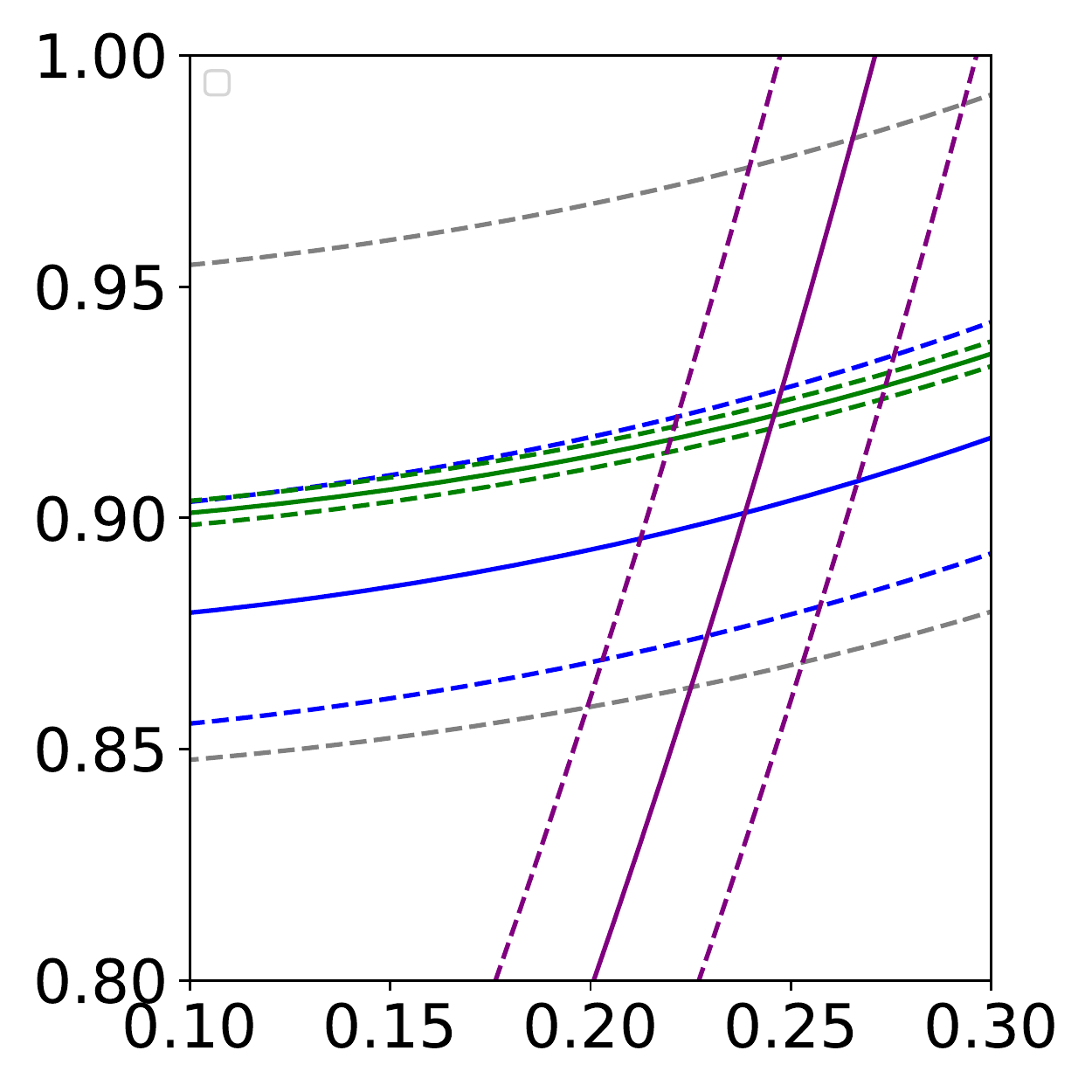}%
    }}
    \end{subfigure}
    \caption{Representation of measurements of three PK parameters in future assuming timing campaigns with FAST until 2039, which is when the mass uncertainties reach kinematic precision. The dashed lines show 1$\sigma$ uncertainties, and dashed grey lines represent the kinematic limit on the $\Pbdot$ measurement coming from distance uncertainty.}
    \label{fig:mass-mass-sim}
  \end{figure}

We created ToAs with \TEMPOTWO\, software for this purpose. Adding these simulated ToAs to our data yields an improvement of about a factor of 6 in the measurement of $\dot{\omega}$, and an improvement of a factor of 4 in the $h_3$ parameter. This implies that we will get highly precise mass estimates and a compelling test of GR with this system in the future. Figure~\ref{fig:mass-mass-sim} plots the estimates of PK parameters from this simulation on the measurements of component masses and inclination in the $\Mp-\Mc$ and $\cosi-\Mc$ diagrams.

\begin{figure}
\centering
        \includegraphics[width=\columnwidth]{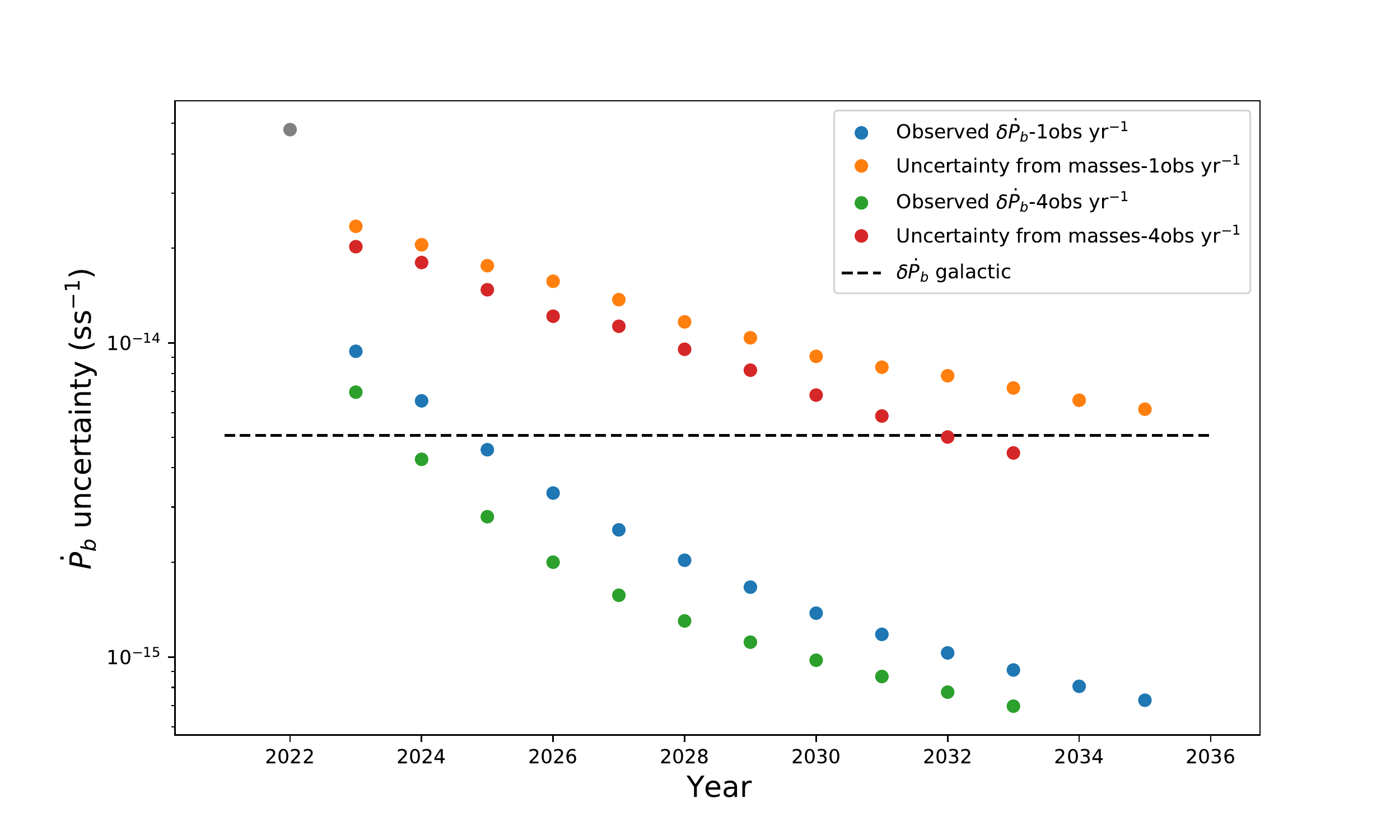}
        \caption{$\Pbdot$ uncertainty variation with simulated orbital campaigns for campaigns once (blue) and four (green) times every year. The precision of $\Pbdot$ predicted by GR as $\dot{\omega}$ uncertainty decreases over the years. This is shown with orange and red points. The grey point shows the current value, and the dashed line represents the limit due to kinematic uncertainty.}
        \label{fig:pbdot-sim}
\end{figure}

Even though the detection significance of all the PK parameters will improve by at least a factor of 4$\sigma$, the improvement in the orbital decay measurement will be much faster (because its uncertainty is proportional to $T^{-5/2}$, where $T$ is the timing baseline).
However, this improvement will soon halt because of the limited precision of the kinematic contributions to the observed $\Pbdot$ (due to the lack of precise distance estimates) of 5.1 $\rm fs \, \rm s^{-1}$. Assuming an orbital campaign is performed with FAST every year, the simulated precision will reach this limit in 2025, as shown in Figure~\ref{fig:pbdot-sim}, requiring only three more orbital campaigns with FAST. 
This precise $\Pbdot$  together with the improved measurement of $h_3$ will yield a very significant improvement in the component masses under the assumption of GR.
%after that, the uncertainty in observed orbital decay will reach the precision limit due to kinematic contribution to orbital decay. 

From then onwards, the quality of the GR test with this system will continue to improve because of the improvements from $\dot{\omega}$ and $h_3$. This will take longer because for these parameters, the uncertainties decrease at a slower rate, $T^{-3/2}$ and $T^{-1/2}$ , respectively. The GR test stops improving when the uncertainty on $\dot{\omega}$ yields a value of $M_{\rm T}$ that is more precise than that derived from $\dot{P}_{b}$ (which is limited by the uncertainty in the estimate of the kinematic contribution to $\Pbdot$). This will only happen after 2032 if four campaigns every year are carried out, and after 2036 if a yearly campaign is carried out (see Figure~\ref{fig:pbdot-sim}). If more sparse observations are assumed, with observations once every two years until 2027 and once every three years, we will reach this limit only by 2039.

Because the measurement of the orbital decay will be limited in the future, we can use this limit to constrain the difference of the effective scalar coupling strengths. The dipolar GW emission contribution to $\Pbdot$ must be smaller than this precision. Thus assuming $\delta\Pbdot^D$ of $5.1\times 10^{-15}$, we obtain $|\alphap-\alphac| < 1.30 \times 10^{-3}$ (68\% C.L.), which is marginally better than the current limit from PSR~J1738+0333, but independent of WD atmosphere models.
Therefore, PSR~J1952+2630 has the potential to provide very stringent radiative tests of certain alternative gravity theories in the near future. 

The analysis based on the measured excess rate of the orbital decay $\Pbdotx$ is straightforward but has some disadvantages. In a specially selected STT the effective coupling strength of a compact object can strongly depend on the mass, especially for an NS ($\alphap$) due to the scalarisation effect \citep{1993PhRvL..70.2220D}. Thus, the importance of the derived limit on the difference of effective scalar coupling strengths $|\alphap-\alphac|$ significantly depends on the masses of a binary system. The large uncertainty in the mass measurements for PSR~J1952+2630 forces us to investigate the variation in $\alphap$ value with a mass. We now discuss the fully consistent analysis in more detail.

\subsection{Constraints on DEF gravity}
\label{sec:DEF}

\begin{figure}
\centering
        \includegraphics[width=\columnwidth]{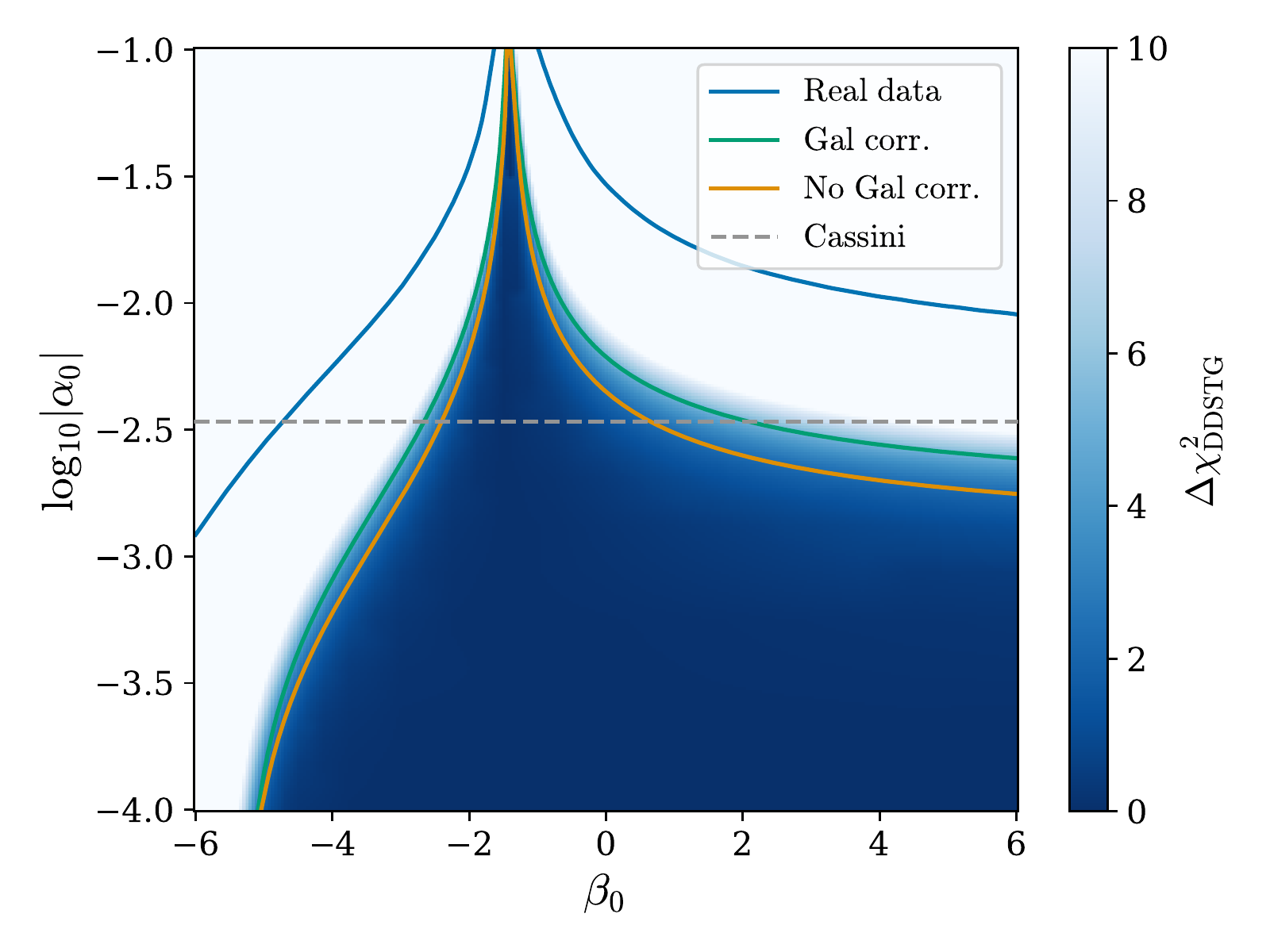}
        \caption{$\Delta \chi^2$ map in DEF gravity space $\{\alpha_0,\beta_0\}$ for the simulated dataset (2023-2032) with four campaigns every year. The tests are performed using the DDSTG timing model and assuming stiff MPA1 EOS. The solid lines correspond to the $68\%$ C.L. limits ($\Delta\chi^2 \simeq 2.28$). The limit with the blue line is based on the existing TOAs. The blue shade and the orange line are the results for the simulated TOAs that do not account for the Galactic uncertainty in $\dot{P}_\mathrm{b,xs}$. The green line shows the corrected limit, with $\dot{P}_\mathrm{b,xs}$ taken at its 1$\sigma$ limit. The dashed grey line is the limit from the Solar System Cassini experiment ($\alpha_0^2 \lesssim 1.15\times10^{-5}$).}
        \label{fig:DDSTG_DEF_limits}
\end{figure}

\begin{figure}
\centering
        \includegraphics[width=\columnwidth]{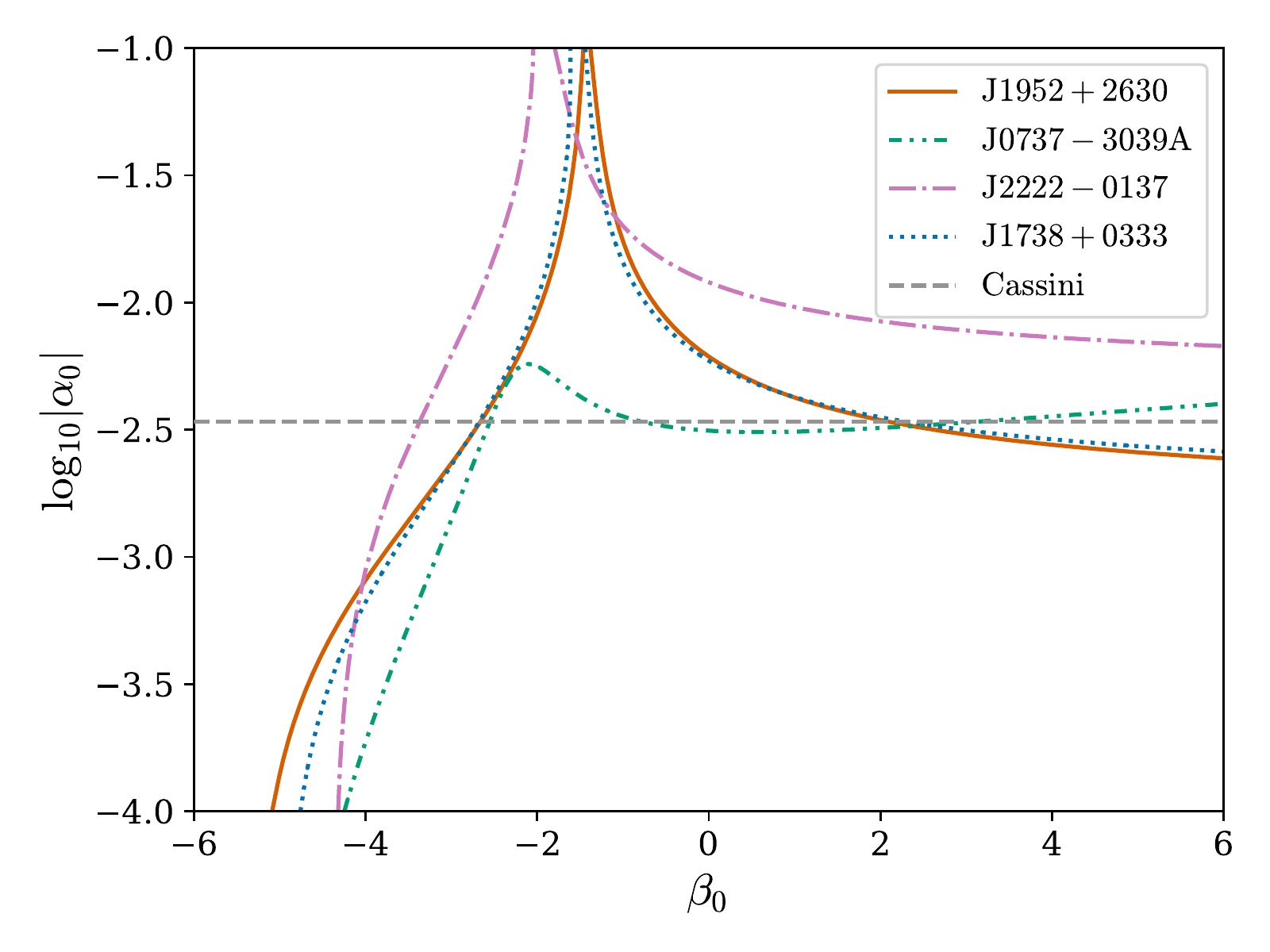}
        \caption{Comparison of existing limits placed on DEF gravity space $\{\alpha_0,\beta_0\}$ by radiative tests from different pulsars to the predictions for J1952+2630. All limits are taken for $68\%$ C.L. and assume MPA1 EOS. The limit for J1952+2630 is predicted by the DDSTG timing model based on the simulated data set for observations from 2023 until 2032 with four campaigns every year. The limits for other pulsars are calculated by the traditional PK method based on the current measured PK parameters published in the literature (see Table~\ref{tab:DEF_pulsars}). We do not include the limit from the triple system PSR J0337+1715 because of its non-radiative origin, despite being more restrictive in the large range of $\beta_0$.}
        \label{fig:DDSTG_DEF_comparison}
\end{figure}

In this section we would like to concentrate on a particular class of STT, the  DEF gravity \citep{1993PhRvL..70.2220D, 1996PhRvD..54.1474D}. DEF gravity is a massless scalar-tensor gravity theory with one long-ranged scalar field $\varphi$ non-minimally coupled to the curvature scalar. The theory is described by two arbitrary parameters $\{\alpha_0, \beta_0\}$ that enter the coupling function that determines how the scalar field $\varphi$ couples to matter. DEF gravity fully recovers GR when both parameters approach zero: $\alpha_0 =0, \beta_0 = 0$. The theory predicts the well-investigated phenomenon of spontaneous scalarisation of NSs. It is a fully nonpertubative effect \citep{1993PhRvL..70.2220D}, resulting in the strong growth of the scalar field in the interiors of the compact object while the scalar field in the exterior space-time stays negligible. Compact objects in DEF gravity obtain special gravitational form-factors (i.e. effective coupling strengths $\alphap, \alphac$, their derivatives with respect to the scalar field, etc.) that enter the observable variables (PK parameters) and can be directly calculated from the structure equations. A very important prediction of DEF gravity that can be tested with binary pulsars is the dipolar contribution to the rate of the orbital period change $\dot{P}_\mathrm{b}^{\mathrm{D}}$. For this paper, we used DEF gravity as a powerful framework with two arbitrary parameters $\{\alpha_0, \beta_0\}$ to perform radiative tests of gravity.

We applied the DDSTG timing model to place limits on DEF gravity (Batrakov et al. in prep). This novel approach is superior to the traditional method based on the measured PK parameters (PK method). The DDSTG model is based on the aforementioned DD model and was developed specifically to constrain scalar-tensor gravity theories. It is implemented in the \TEMPO \, software and uses the theoretical predictions for PK parameters in DEF gravity internally. PK parameters in DEF gravity depend on two masses of the pulsar and the companion, their gravitational form-factors, and the choice of the equation of state (EOS). The model is supplied with pre-calculated grids of NSs gravitational form-factors for a set of different EOS. For a WD companion, we assume a weak-field approximation $\alphac=\alpha_0$ , which is sufficient due to the small compactness of a WD.

The DDSTG model fits two masses of the objects along with Keplerian, spin, and astrometric parameters directly to the timing data without any intermediate steps. The timing data can show strong correlations that are hard to account for using the traditional PK method \citep{2019CQGra..36v5009A}. In contrast, the direct fit of the theory to the data naturally accounts for all possible correlations between PK and other parameters and exploits weakly measured relativistic effects. 
%TG: Removed para below.

%The timing data can show strong correlations which are hard to account for using the traditional PK method \citep{2019CQGra..36v5009A}. The problem is even worse if the relation is nonlinear (i.e. the Shapiro delay parameters, $r \sim s$) or the effect is present in the data but the respected parameter cannot be measured. The DDSTG model also accounts for the parameters which cannot be predicted from the theory but affect the observables, for example the change of the projected semi-major axis $\dot{x}$ correlates with the time dilation parameter $\gamma$ in PSR J1141-6545 \citep{2019MNRAS.490.3860R}. In that system $\dot{x}$ is the result of the spin-orbit coupling due to the fast rotating WD and the arising precession can not be calculated because the spin of the WD in unknown.

In Figure~\ref{fig:DDSTG_DEF_limits} we present limits on DEF gravity obtained by the DDSTG model for the PSR J1952+2630. We compare the limits calculated for the existing data set ($1475$ TOAs) and the simulated one ($18752$ TOAs from $2023-2032$ with four campaigns per year). For the present investigation, we assumed the stiff EOS MPA1 \citep{1987PhLB..199..469M} in the piecewise polytropic approximation \citep{2009PhRvD..79l4032R} and used the corresponding grid of gravitational form-factors. MPA1 has a relatively high maximum mass of the NS $M_\mathrm{max} = 2.461 M_\odot$. Stiff equations of state generally produce more conservative limits on DEF gravity for most of the $\{\alpha_0, \beta_0\}$ plane compared to soft equations of state. For a selected pair of $\{\alpha_0, \beta_0\}$ parameters, \TEMPO \, fits the best $\chi^2$ value. The minimum $\chi^2_\mathrm{min}$ in $\{\alpha_0, \beta_0\}$ domain corresponds to the statistically preferred theory parameters. The obtained $\chi^2_\mathrm{min}$ value statistically agrees with the GR value $\chi^2_\mathrm{GR}$ for $\alpha_0 =0, \beta_0 = 0$. The shifted quantity $\Delta\chi^2 = \chi^2 - \chi^2_\mathrm{min}$ has the $\chi^2$ statistics with 2 degrees of freedom and was used to perform statistical tests. We derived contours of $\Delta\chi^2$ to place limits on $\{\alpha_0, \beta_0\}$ within the desired confidence level limit (we used $68\%$ C.L., which corresponds to $\Delta \chi^2 \simeq 2.28$). 

\TEMPO\, does not allow us to internally account for the uncertainty in $\dot{P}_\mathrm{b,x}$ due to the Galactic and Shklovskii contributions. The value $\dot{P}_\mathrm{b,x}$ was therefore set fixed for each run of \TEMPO. Ideally, a dense grid of $\chi^2$ over $\alpha_0 - \beta_0 - \dot{P}_\mathrm{b,x}$ parameters is calculated and then marginalised over $\dot{P}_\mathrm{b,x}$ , with the prior corresponding to the uncertainty of the $\dot{P}_\mathrm{b,x}$ term. However, this approach is very intensive computationally. In Figure~\ref{fig:DDSTG_DEF_limits} we present the contours for two fixed $\dot{P}_\mathrm{b,x}$ values. One contour is for the calculated prediction of $1.8$ fs s$^{-1}$ , and the second is shifted by the uncertainty $1.8 + 5.1 = 6.9 $ fs s$^{-1}$. The $\Delta\chi^2$ contour with the $\dot{P}_\mathrm{b,x}$ value shifted by $1\,\sigma$ uncertainty gives a slightly conservative but still reasonable limit on DEF gravity parameters. The more accurate account for the $\dot{P}_\mathrm{b,x}$ uncertainty would give slightly better results in between the shifted and non-shifted contours.

To place limits on DEF gravity parameters $\{\alpha_0, \beta_0\}$ , the pulsar must have at least three measured PK parameters (or other parameters depending on masses, e.g. the mass ratio $q$). All radiative tests strongly depend on the precision in the $\dot{P}_\mathrm{b}$ measurement. PSR J1952+2630 shows $\dot{P}_\mathrm{b}, \dot{\omega}$, and $h_3$ measurements at present (see Table~\ref{tab:DEF_pulsars}), and $\varsigma$ prediction in the future. Two important limiting factors that affect the test of the DEF gravity in this system are a) the precision of $\dot{\omega}$, and b) the mass of the pulsar $\Mp$. The spontaneous scalarisation of the NS, which predicts the strong dipolar contribution $\delta\Pbdot^D$ term, occurs in the nonlinear area with $\beta_0 \lesssim -4.4$ and only for relatively massive NS, $M_\mathrm{p} \gtrsim 1.5 \rm M_\odot$, for instance, PSR J2222$-$0137 \citep{2021A&A...654A..16G}. The current measurement of $M_\mathrm{p} \sim 1.2 \rm M_\odot$ for PSR J1952+2630 is relatively small and thus does not place a restrictive limit in this region. However, the error bars on the masses are large, so that the constraints from this pulsar can become restrictive in the future if $M_\mathrm{p}$ is found to be higher. 

Figure~\ref{fig:DDSTG_DEF_comparison} shows a comparison between the limits that PSR J1952+2630 will present in the future (assuming four orbital campaigns per year from 2023-2032) and the current best constraints placed by other systems on the $\{\alpha_0, \beta_0\}$ plane. We include systems that are powerful in terms of radiative tests (see Table~\ref{tab:DEF_pulsars}). PSR J1952+2630 can place restrictive limits comparable to the current limits from PSR J1738+0333 ($M_\mathrm{p} \sim 1.46 M_\odot$ and $M_\mathrm{c} \sim 0.18 M_\odot$)  for the positive $\beta_0 > 2$ region. The latter system has been dominant this region so far \citep{2012MNRAS.423.3328F}, but since the precision on its component masses is limited \citep{2012MNRAS.423.3316A}, there is little scope for further improvements. This will not be the case for PSR J1952+2630 as the constraints on mass measurements will continue to improve with precise $\dot{\omega}$ values in future. As discussed above, the limiting constraints for this pulsar will eventually come from the poor estimate of the kinematic contributions due to large uncertainties in its distance measurement. Conversely, pulsar binaries such as PSR J2222$-$0137 and the double pulsar J0737$-$3039A have the potential to provide further restrictive constraints on the radiative tests in the future.
%The system PSR J2222$-$0137 (comprising a massive NS, $M_\mathrm{p} \sim 1.82 M_\odot$ and a massive WD, $M_\mathrm{c} \sim 1.31 M_\odot$) currently puts the best limits on the scalarization region with large negative $\beta_0$ and small $\alpha_0$ \citep{2021A&A...654A..16G}, however, it can also become powerful for the positive $\beta_0$ region in the future (for instance see Batrakov et al. in prep.). The Double Pulsar J0737$-$3039A ($M_\mathrm{p} \sim 1.34 M_\odot$ and $M_\mathrm{c} \sim 1.25 M_\odot$) consists of two neutron stars and is restrictive in a intermediate area near $\beta_0 \sim 0$ and becomes less restrictive in positive $\beta_0 > 2$ region. 
The discussion above shows that radiative tests in terms of DEF gravity framework are characterised not only by the constraint on the $|\alphap-\alphac|$ parameter, but also by the properties of a binary system such as the companion masses.

\begin{table}
    \caption{Binary pulsar systems used to compare constraints on the DEF theory in Figure~\ref{fig:DDSTG_DEF_comparison}. The measured PK parameters published in the literature along with other information are used for calculating limits. $\Mc$ and the mass ratio $q$ for J1738+0333 are measured from the optical observations of the companion. The value in brackets is expected to be measured in the future.}
    \begin{center}
        \begin{tabular}{lcc}
            \hline \hline
            Name & Observed parameters & reference \\ 
            \hline
            J1952+2630 & $\dot{\omega}, \dot{P}_\mathrm{b}, h_3, (\varsigma) $ & this work \\
            J0737$-$3039A & $\dot{\omega}, \gamma, \dot{P}_\mathrm{b}, r, s, q$ & \text{\cite{2021PhRvX..11d1050K}} \\
            J2222$-$0137 & $\dot{\omega}, \dot{P}_\mathrm{b}, h_3, \varsigma $ & \text{\cite{2021A&A...654A..16G}} \\
            J1738+0333 & $\dot{P}_\mathrm{b}, M_\mathrm{c}, q$ & \text{\cite{2012MNRAS.423.3328F}} \\
            \hline
            \hline
        \end{tabular}
    \end{center}
    \label{tab:DEF_pulsars}
\end{table}
%TG: mass ratio already defined in section 6.

\subsection{Additional orbital parameters}

 We have measured no additional PK parameters nor any other constraints on the orbital geometry of the PSR~J1952+2630 system from our current timing data. The simulations we made can be used to determine whether such measurements might be possible.

The most likely detection is that of the second Shapiro delay parameter ($\varsigma$ in our chosen parametrisation). This is less secure than the detection of $h_3$, as it depends more strongly on the largely unknown orbital inclination of the system (see Figure~\ref{fig:mp-mc-cosi-planes}). In our simulations, which assumed an inclination of 74$^\circ$(close to the peak of probability in that figure), we can obtain a significant detection of this parameter, about 14$\sigma$. However, for lower inclinations, which are still possible, the significance of this detection will be much lower.

Table \ref{tab:timing_solution_J1952} shows  that the value of the relativistic $\gamma$ (which quantifies the slowdown of the pulsar at periastron relative to apastron caused by a combination of special relativistic time dilation and general relativistic gravitational redshift) expected for the most likely mass combination is very low, about $0.2 \, \mu$s. The reason is clearly the very low orbital eccentricity of this system. Measuring this effect in our simulations, we obtain an uncertainty of 5$\,\mu$s, which is certainly not enough for the detection of such a small $\gamma$.

Another effect we investigated is the secular change of the projected semi-major axis, $\dot{x}$. An analysis of the possible different contributions to this effect leads us to conclude that the largest term  by far is the change in viewing angle on the binary caused by its proper motion \citep{1996ApJ...467L..93K}. For an inclination of $74^\circ$ , we obtain $|\dot{x}| < \, 7 \times 10^{-16}$ lt-s s$^{-1}$. When we try to measure this effect from our simulated data, we obtain an uncertainty of $3.7 \times 10^{-15}$ lt-s s$^{-1}$, which is certainly not sufficient for the detection of this effect either.

We note, in addition, that for a timing baseline that is much shorter than the full precession timescale ($360^\circ / \dot{\omega} \sim 200$ yr, in this case), the effect of $\gamma$ on the timing cannot be clearly distinguished from that of $\dot{x}$ \citep{2019MNRAS.490.3860R}. Thus, a detection of $\gamma$ will require much longer baselines than the one we have simulated.

\section{Summary and conclusions}
\label{sec:discussion}
We presented the results from the continued timing of PSR~J1952+2630 with the Arecibo telescope. We showed the polarimetric profile of this pulsar and measured RM of $-145\pm0.15 \, \rm rad \, m^{-2}$. However, a fit for PA swing was unable to yield a good measure of its inclination. We showed improved precision on astrometric, orbital, and kinematic parameters, with an improvement of an order of magnitude in the proper motion of 5.89(23) mas yr$^{-1}$.

For the first time, we presented the measurement of three PK parameters: $\dot{\omega}$, $\Pbdot$, and the orthometric amplitude of the Shapiro delay, $h_3$. Assuming GR, these yield $M_{\rm p}$ to 1.24(26) \msun \, and $\Mc$ to 0.95(12) \msun, indicating that the companion likely is a CO WD formed from a He star. Because of the uncertainty of $\Mc$, the possibility of an ONeMg WD is less likely, but still cannot be ruled out.

This system has some promise as a gravitational laboratory.
First, with our improved proper motion estimates, we calculated the kinematic contributions to the observed orbital decay due to Galactic accelerations and Shklovskii effect. The uncertainty of these contributions (due to the limited distance measurement) represents the ultimate limit on the improvement in the measurement of the orbital decay of this system.

Despite this limitation, the system has the potential of placing stringent constraints on the 
DEF gravity parameters $\{\alpha_0, \beta_0\}$ , similar to PSR~J1738+0333, which is currently the best radiative test in the $\beta_0 > 2$ region, but with the advantage that the masses are determined from timing measurements and are therefore independent of WD spectroscopy and its interpretation with WD atmosphere models.
Our simulations show that such a test is feasible within the near future:
much improved masses will be known by 2025, and a test comparable to that of PSR J1738+0333 by 2032 if four short orbital campaigns are performed every year with FAST.

The early knowledge of the component masses will be important not only to elucidate the nature of the WD companion, but also for a better understanding of the possibilities of this system for tests of gravity theories. In particular, if $\Mp > 1.5 \, \rm M_{\odot}$ (which is still allowed by the current uncertainties), then measurements with this system might provide not only excellent constraints on $\alpha_0$, but also introduce constraints on the phenomenon of spontaneous scalarisation that are even tighter than current limits on this phenomenon (see Figure 2 of \citealt{2022CQGra..39kLT01Z}).

One final remark about the timing of this system. At first sight, its low eccentricity (only $4.2 \times 10^{-5}$) and unremarkable Shapiro delay (barely detectable) do not suggest that it is a promising gravity laboratory. All relativistic effects measured in this system to date are small and difficult to measure, resulting in PK parameters with large uncertainties. However, a dedicated study has revealed that an apparently unremarkable system might still have much to teach us. This shows that detailed studies of apparently unpromising systems can occasionally reveal real jewels.

\section*{Acknowledgements}
We thank Norbert Wex for useful discussions and suggestions. TG thanks Alessandro Ridolfi for help with data analysis of polarization data. The Arecibo Observatory is a facility of the National Science Foundation operated under cooperative agreement by the University of Central Florida and in alliance with Universidad Ana G. Mendez, and Yang Enterprises, Inc. . This work made use of the data from FAST (Five-hundred-meter Aperture Spherical radio Telescope). FAST is a Chinese national mega-science facility, operated by National Astronomical Observatories, Chinese Academy of Sciences. EP supported by the H2020 ERC Consolidator Grant "MAGNESIA" under grant agreement No. 817661 and National Spanish grant PGC2018-095512-BI00.

%%%%%%%%%%%%%%%%%%%%%%%%%%%%%%%%%%%%%%%%%%%%%%%%%%
%\section*{Data Availability}


\bibliographystyle{mnras}
\bibliography{44699corr} % if your bibtex file is called example.bib

%\begin{thebibliography}{99}
%\end{thebibliography}

%%%%%%%%%%%%%%%%%%%%%%%%%%%%%%%%%%%%%%%%%%%%%%%%%%

%%%%%%%%%%%%%%%%% APPENDICES %%%%%%%%%%%%%%%%%%%%%

%\appendix



%%%%%%%%%%%%%%%%%%%%%%%%%%%%%%%%%%%%%%%%%%%%%%%%%%

% Don't change these lines
%\bsp   % typesetting comment
\label{lastpage}
\end{document}